\documentclass[3p, sort&compress, pdflatex]{elsarticle}

\usepackage{amssymb, amsmath}

\usepackage{graphicx}
\usepackage{subfig}
\usepackage[ruled, linesnumbered]{algorithm2e}
\usepackage{hyperref}
\usepackage{todonotes}

\usepackage{enumitem}

\newcommand{\bs}[1]{\boldsymbol{#1}}
\newcommand{\mdif}[1]{\dfrac{\mathrm{D} #1}{\mathrm{D}t}}
\newcommand{\pdif}[2]{\dfrac{\partial #1}{\partial #2}}

\begin{document}

\begin{frontmatter}

\title{Computational co-design of structure and feedback controller for locomoting soft robots}

\author{Yuki Sato}
\author{Changyoung Yuhn}
\author{Hiroki Kobayashi}
\author{Atsushi Kawamoto}
\author{Tsuyoshi Nomura \corref{cor1}}
\ead{nomu2@mosk.tytlabs.co.jp}
\cortext[cor1]{Corresponding author.}

\address{Toyota Central R\&D Labs., Inc., 1-4-14, Koraku, Bunkyo-ku, Tokyo, 1120004, Japan}

\begin{abstract}
Soft robots have gained significant attention due to their flexibility and safety, particularly in human-centric applications. 
The co-design of structure and controller in soft robotics has presented a longstanding challenge owing to the complexity of the dynamics involved. 
Despite some pioneering work dealing with the co-design of soft robot structures and actuation, design freedom has been limited by stochastic design search approaches.
This study proposes the simultaneous optimization of structure and controller for soft robots in locomotion tasks, integrating topology optimization-based structural design with neural network-based feedback controller design.
Here, the feedback controller receives information about the surrounding terrain and outputs actuation signals that induce the expansion and contraction of the material. 
We formulate the simultaneous optimization problem under uncertainty in terrains and construct an optimization algorithm that utilizes automatic differentiation within topology optimization and neural networks.
We present numerical experiments to demonstrate the validity and effectiveness of our proposed method.
\end{abstract}


\end{frontmatter}

\section{Introduction} \label{sec:intro}

Soft robotics has attracted a growing interest, attributed to its inherent flexibility, versatility, and safety, particularly in human-centric applications.
It has wide applications, including exploration tasks, object manipulators, medical usages, and wearable devices~\cite{laschi2016soft, lee2017soft, yasa2023overview}.
For designing functional soft robots, the co-design of both structure and controller has presented a persistent challenge due to the complex dynamics, including the large deformation and interaction with their surrounding environment.
Cheney et al.~\cite{cheney2014unshackling} performed a pioneering study for co-designing the structure and actuation of soft robots for locomotion tasks based on the compositional pattern-producing networks (CPPN).
Van Diepen and Shea~\cite{van2022co} presented the co-designing of structures and actuation of soft robots where the structures and actuation were optimized by simulated annealing.
Bhatia et al.~\cite{bhatia2021evolution} proposed a co-design method of soft robots using the CPPN for structural designs and reinforcement learning for the controller designs.
They applied it for various tasks, including locomotion tasks and object manipulation tasks.
Despite pioneering efforts exploring the co-design for soft robots, the degrees of design freedom for structures were relatively limited, primarily due to the reliance on stochastic design exploration methods.

As a structural optimization method with high degrees of design freedom, topology optimization~\cite{bendsoe1988generating} has gained much attention over decades.
The major challenges to applying it for the structural design of soft robots are handling large deformations and contacts in forward analysis, usually performed by the finite element method (FEM)~\cite{zienkiewicz2000finite, hughes2012finite}.
While topology optimization has been originally studied in designing stiff structures~\cite{eschenauer2001topology, deaton2014survey}, there have been extensive studies that considered material and geometrical non-linearity for large deformations in the literature~\cite{buhl2000stiffness, jung2004topology, zhang2021topology, han2017topology}.
Several studies also handled contacts of materials with themselves or other materials in addition to large deformations~\cite{fernandez2020topology, frederiksen2023topology}.

Material point methods (MPMs)~\cite{stomakhin2013material, stomakhin2014augmented, jiang2016material, hu2018moving} are potential alternatives for performing forward analysis involving large deformations and contacts of materials with other materials or boundaries.
Compared with FEMs, MPMs can easily handle large deformations and contacts owing to their hybrid Lagrangian--Eulerian approach.
MPMs have been incorporated in topology optimization~\cite{li2021lagrangian} for elastic materials with large static deformations.
In more recent studies, we have introduced MPMs in topology optimization for locomoting soft robots~\cite{sato2023topology}.
Furthermore, we have proposed \textit{4D topology optimization}~\cite{yuhn20234d}, which simultaneously optimizes the structure, actuator layout, and actuation pulse sequences of a soft robot by representing them as multi-index density variables in four dimensions (i.e., space and time).
The key technique of these studies is the incorporation of MPMs with automatic differentiation, which assists in computing gradients involving forward analysis even for highly complex dynamics including large deformations and contacts.
The experimental demonstration has also been reported, which enhanced the effectiveness of this approach~\cite{kobayashi2023computational}.
Previous efforts, however, have focused on optimizing actuation signals for a predetermined environment rather than creating a controller system that is adaptable to environmental changes.
Hence, the optimized soft robots were operated by feedforward controllers and may be less robust under uncertainties in environments.

To address the environmental uncertainties, we focus on the machine learning (ML)-based controllers for soft robots given that ML techniques have been widely used for soft robot control in the literature~\cite{george2018control, chin2020machine, kim2021review}.
Indeed, previous works for the co-design of soft robots mentioned above successfully dealt with the soft robot control based on the ML-based techniques~\cite{cheney2014unshackling, van2022co, bhatia2021evolution}.
However, the effectiveness of such ML techniques for topology optimization is currently controversial, especially for direct design models without using physical properties~\cite{woldseth2022use} although there are several previous efforts~\cite{abueidda2020topology, zheng2021generating, zheng2023unifying}.
Therefore, it is important to consider how to integrate ML-based techniques into the well-established gradient-based topology optimization for structural designs.

\begin{figure}[t]
    \centering
    \includegraphics[width=1.0\textwidth]{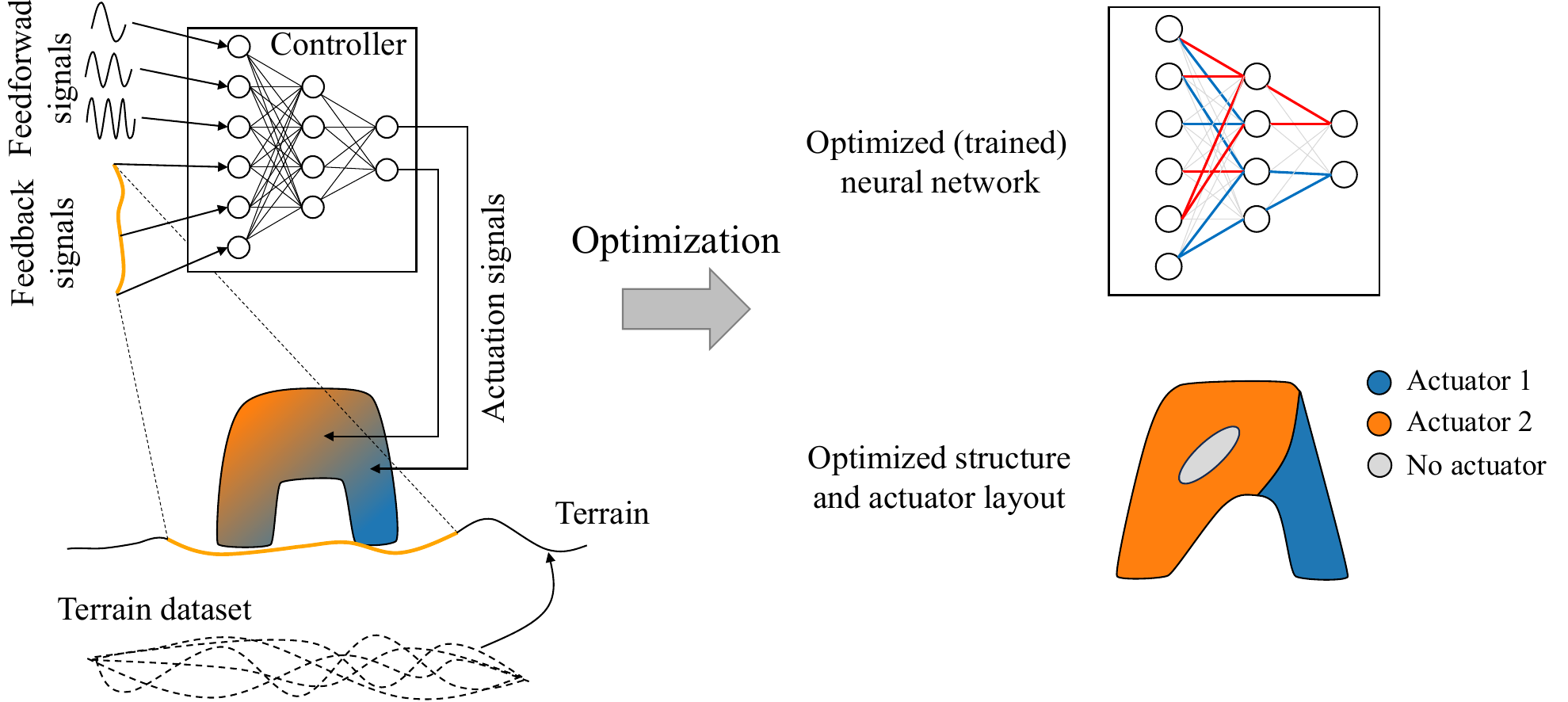}
    \caption{Conceptual diagram of our proposed method.}
    \label{fig:concept}
\end{figure}

In this study, we propose simultaneous optimization of both structure and feedback controller for soft robots involving locomotion tasks, integrating gradient based-topology optimization for structural designs with neural network-based controller designs.
Meanwhile, we use neural networks for feedback controller designs~\cite{jiang2017brief, chin2020machine}, specifically a multilayer perceptron~\cite{du2021diffpd}.
The neural network-based controller generates actuation signals based on terrain features surrounding the soft robots, inducing the expansion and contraction of the soft bodies.
We also optimize the actuator layout based on our previous method~\cite{yuhn20234d}.
Figure~\ref{fig:concept} illustrates the concept of our proposed method.

The contributions of our current work are as follows:
\begin{itemize}
    \item We formulate the co-design problem of a structure and controller of a soft robot for locomotion tasks. We set the design space of structure and actuator layouts based on the topology optimization approach while we model the feedback controller using neural networks.
    \item We construct an optimization algorithm for co-designing soft robots that move over various terrains. To account for the uncertainty of the terrain, our algorithm prepares random terrain datasets and trains neural networks simultaneously with the structure and actuator layout.
    \item We demonstrate the effectiveness of our proposed method by providing forward analysis results using test terrain datasets that are not used in the training (optimization) process.
\end{itemize}

The remainder of this paper is organized as follows.
In Section~\ref{sec:formulation}, we explain the governing equations for forward problems, which include the material and actuation models.
We also briefly discuss the discretization of the governing equation using MPMs.
In Section~\ref{sec:optimization}, we formulate the optimization problem for co-designing soft robots.
We also explain the dataset preparation of varying terrains and construct the optimization algorithm of the structures, actuator layouts, and feedback controllers of soft robots.
In Section~\ref{sec:experiments}, we provide the numerical experiments to demonstrate the validity and the effectiness of our proposed method.
Finally, we conclude this study in Section~\ref{sec:conclusion}.

\section{Forward problem} \label{sec:formulation}

\subsection{Governing equations} \label{sec:gov}
Let $\Omega(t)$ denote a material domain at time $t$.
The conservation laws of mass and momentum in the domain $\Omega(t)$ are given as
\begin{align}
    & \mdif{\rho(t, \bs{x})} + \rho(t, \bs{x}) \nabla \cdot \bs{v}(t, \bs{x}) = 0, \label{eq:mass} \\
    & \rho(t, \bs{x}) \mdif{\bs{v}(t, \bs{x})} = \nabla \cdot \bs{\sigma}(t, \bs{x}) + \bs{f}(t, \bs{x}), \label{eq:momentum}
\end{align}
where $\bs{x}$ is the spatial coordinate, $\rho$ is the material density, $\bs{v}$ is the velocity, $\bs{\sigma}$ is the Cauchy stress tensor, and $\bs{f}$ is the external force vector.
The operator $\mathrm{D} / \mathrm{D} t$ denotes the material derivative.
Hereinafter, we will omit the arguments $(t, \bs{x})$ if apparent from the context.
Assuming the hyperelastic material governed by the neo-Hookean material constitution, we derive the Cauchy stress tensor via the potential energy $\Psi$ and the deformation gradient tensor $\bs{F}$, as follows:
\begin{align}
    & \bs{\sigma} = \frac{1}{J} \pdif{\Psi}{\bs{F}} \bs{F}^\top + \eta \left( \nabla \bs{v} + \nabla \bs{v}^\top - \frac{2}{3} \left(\nabla \cdot \bs{v} \right) \bs{I} \right), \label{eq:Cauchy} \\
    & \Psi(\bs{F}) := \dfrac{\mu}{2} \left( \mathrm{tr}(\bs{F}^\top \bs{F}) - d \right) - \mu \log (J) + \dfrac{\lambda}{2} \log^2 (J),
\end{align}
where $J:=\mathrm{det}(\bs{F})$, $\lambda$ and $\mu$ are the Lam\'{e} constants, $d$ is the spatial dimension, $\eta$ is a damping parameter and $\bs{I}$ is the identity matrix with the size of $d \times d$.
The second term of the Cauchy stress is the viscosity term for numerical damping which is added to avoid numerical instabilities of high-frequency oscillations.
For the external force, we applied the gravity and the self-actuating force as
\begin{align}
    & \bs{f}(t, \bs{x}) = - \nabla \cdot \bs{\sigma}^\text{act}(t, \bs{x}) + \rho(t, \bs{x}) \bs{g}, \label{eq:force} \\
    & \bs{\sigma}^\text{act}(t, \bs{x}) := - a(t, \bs{x}) \bs{F}(t, \bs{x}) \bs{S} \bs{F}(t, \bs{x})^\top,
\end{align}
where $a(t, \bs{x})$ is the actuation signal, $\bs{g}$ is the gravity acceleration and $\bs{S}$ is the second Piola--Kirchhoff stress tensor prescribing the deformation caused by the actuation.
In the present study, we set $\bs{S}:= \bs{I}$.
Since we consider the locomoting soft robots, we have to consider boundary conditions that describe the interaction of the soft robots with the boundaries of the environment, e.g., ground and wall.
In the present study, we consider the no-slip ground that imposes
\begin{align}
    \bs{v}(t, \bs{x}) = \bs{0} \text{ for } \bs{x} \text{ on the ground.} \label{eq:bnd}
\end{align}

\subsection{Discretization}
To solve the governing equations in Eqs.~\eqref{eq:mass} and \eqref{eq:momentum} with the boundary condition in Eq.~\eqref{eq:bnd}, we used the moving least square material point method (MLS-MPM)~\cite{hu2018moving}, which is a variant of material point methods (MPMs)~\cite{stomakhin2013material, stomakhin2014augmented, jiang2016material}, a Lagrangian-Eulerian hybrid method.
MPMs discretize the material domain into particles and take into account the couples of particles using a background grid.
More precisely, the numerical simulation process of MPMs is the iteration of the following steps:
\begin{enumerate}[label=\textbf{(\arabic*)}]
    \item \textbf{Particle to grid} \\
    Transfer the mass and momentum of particles to nodes of a background grid.
    \item \textbf{Update velocities} \\
    Calculate the velocities on each node of the grid by transferred mass and momentum. Update the velocity on the grid considering the external forces and the boundary conditions.
    \item \textbf{Grid to particle} \\
    Transfer the updated velocity on nodes of the grid to each particle and update positions of each particle.
\end{enumerate}
The boundary condition in Eq.~\eqref{eq:bnd} is imposed in the second step by replacing the grid velocities on the boundary, as follows:
\begin{align} \label{eq:bnd_mpm}
    \text{For } m \in \mathcal{B}, \quad \bs{v}_m \leftarrow \begin{cases}
        \bs{0} & \text{ if } \bs{v}_m \cdot \bs{n}_m < 0 \\
        \bs{v}_m & \text{ otherwise},
    \end{cases}
\end{align}
where $\mathcal{B}$ is the index set of grid nodes on the boundary, $\bs{v}_m$ is the velocity on the $m$-th node of the grid and $\bs{n}_m$ is the normal vector at the $m$-th node pointing outward to the boundary.
This boundary condition corresponds to the no-slip boundary condition and can be easily extended more general boundary condition considering the Coulomb friction~\cite{hu2018moving}.
See the reference~\cite{hu2018moving} for more detailed implementation.

\section{Optimization problem} \label{sec:optimization}

Let us introduce a fixed design domain $\Omega_\text{D}$ that includes the material domain at an initial time $t_0$, i.e., $\Omega(t_0) \subset \Omega_\text{D}$.
We now describe how to optimize the structure, actuator layout, and controller of a soft body.

\subsection{Topology optimization}
We formulate the structural optimization problem of the soft body based on the density-based topology optimization approach~\cite{bendsoe1989optimal, bendsoe2003topology}.
Let $\gamma_{i} \in [0, 1]$ be the fictitious material density for the $i$-th particle where $\gamma_i=1$ for solid particles, $\gamma_i=0$ for void particles and $0 < \gamma_i < 1$ for intermediate particles.
We interpolate the material properties, as follows:
\begin{align}
    & \rho_{i} = \bar{\rho} ((1-\varepsilon) \gamma_{i} + \varepsilon), \\
    & \lambda_{i} = \bar{\lambda} ((1-\varepsilon) \gamma_{i} + \varepsilon), \\
    & \mu_{i} = \bar{\mu} ((1-\varepsilon) \gamma_{i} + \varepsilon),
\end{align}
where $\bar{\rho}$, $\bar{\lambda}$ and $\bar{\mu}$ are the density and Lam\'{e} constants of the solid material, respectively, and $\varepsilon$ is the small constant to avoid numerical instabilities.
We set $\varepsilon = 10^{-5}$ in this study.

To ensure the smoothness of the structures, we apply the particle-wise density filtering which average the fictitious material density of each particle using those of neighboring particles~\cite{yuhn20234d}.
Let $\bs{\phi}:=[\phi_{1}, \cdots, \phi_{N^\text{par}}]$ denote the array of design variables whose component $\phi_{i}$ determines the fictitious material density $\gamma_{i}$ via density filtering and Heaviside projection.
First, we calculate filtered variables $\tilde{\phi}_i$ for each particle, as follows:
\begin{align}
    \tilde{\phi}_i = \frac{\sum_{i'=1}^{N^\mathrm{par}} w(\| \bs{x}_i - \bs{x}_{i'} \|) \phi_{i'}}{\sum_{i'=1}^{N^\mathrm{par}} w(\| \bs{x}_i - \bs{x}_{i'} \|)}
\end{align}
where $\bs{x}_i$ is the position of the $i$-th particle and $w$ is the weighting function defined as
\begin{align} \label{eq:weight_function}
    w(r) := \left( 1 - \frac{ \mathrm{min} (r, R) }{R} \right)^{p},
\end{align}
where $R$ is a filter radius and $p$ is a parameter for determining the decay of the weight.
Applying the Heaviside projection using the sigmoid function, we obtain
\begin{align}
    \gamma_{i} = \frac{1}{1 + \exp\left(-\beta^\mathrm{to} \tilde{\phi}_{i}\right)},
\end{align}
where $\beta^\mathrm{to}$ is a parameter that determines the curvature of the sigmoid function.

\subsection{Actuator layout optimization}
Next, we formulate the layout optimization problem that determines the self-actuating sub-domains in the material domain of a soft body, which is based on our previous work~\cite{yuhn20234d}.
Let $\xi_{ij} \in [0, 1]$ be the continuously relaxed two-index variable for assigning the $j$-th actuation signal to the $i$-th particle.
We define the actuation signal, as follows:
\begin{align}
    a_{i}(t) = ((1-\varepsilon) \gamma_{i}^{3} + \varepsilon) \sum_{j=1}^{N^\text{act}+1} \xi_{ij} \bar{a}_{j}(t),
\end{align}
where $\bar{a}_{j \leq N^\mathrm{act}}$ is the $j$-th actuation signal and $\bar{a}_{N^\mathrm{act} + 1}=0$ for representing no actuation.
The coefficient $((1-\varepsilon) \gamma_{i}^{3} + \varepsilon)$ attenuates the actuation strength on void particles, where the power of $\gamma_i$, 3, functions as a penalty to intermediate values of $\gamma_i$ similar to the solid isotropic material with penalization (SIMP) method~\cite{bendsoe1989optimal, bendsoe1999material}.
To ensure the smoothness of the layout as well as the structure, we use the particle-wise density filtering technique.
Let $\bs{\psi}:=[\psi_{11}, \cdots, \psi_{N^\text{par} N^\text{act}+1}]$ denote the array of design variables whose component $\psi_{ij}$ determines the relaxed multi-index variable $\xi_{ij}$ via density filtering and softmax projection.
First, we calculate filtered variable $\tilde{\psi}_{ij}$ for each particle, as follows:
\begin{align}
    \tilde{\psi}_{ij} = \frac{\sum_{i'=1}^{N^\mathrm{par}} w(\| \bs{x}_i - \bs{x}_{i'} \|) \psi_{i'j}}{\sum_{i'=1}^{N^\mathrm{par}} w(\| \bs{x}_i - \bs{x}_{i'} \|)},
\end{align}
where $w$ is defined in Eq.~\eqref{eq:weight_function}.
Applying the softmax function to the filtered variables, we obtain
\begin{align}
    \xi_{ij} = \frac{\exp\left( \beta^\mathrm{lay} \tilde{\psi}_{ij} \right) }{\sum_{j'=1}^{N^\text{act}+1} \exp\left( \beta^\mathrm{lay} \tilde{\psi}_{ij'} \right)},
\end{align}
where $\beta^\mathrm{lay}$ is a parameter for controlling the curvature of the softmax function. 

\subsection{Controller optimization} \label{sec:controller}
\begin{figure}[t]
    \centering
    \includegraphics[width=0.5\textwidth]{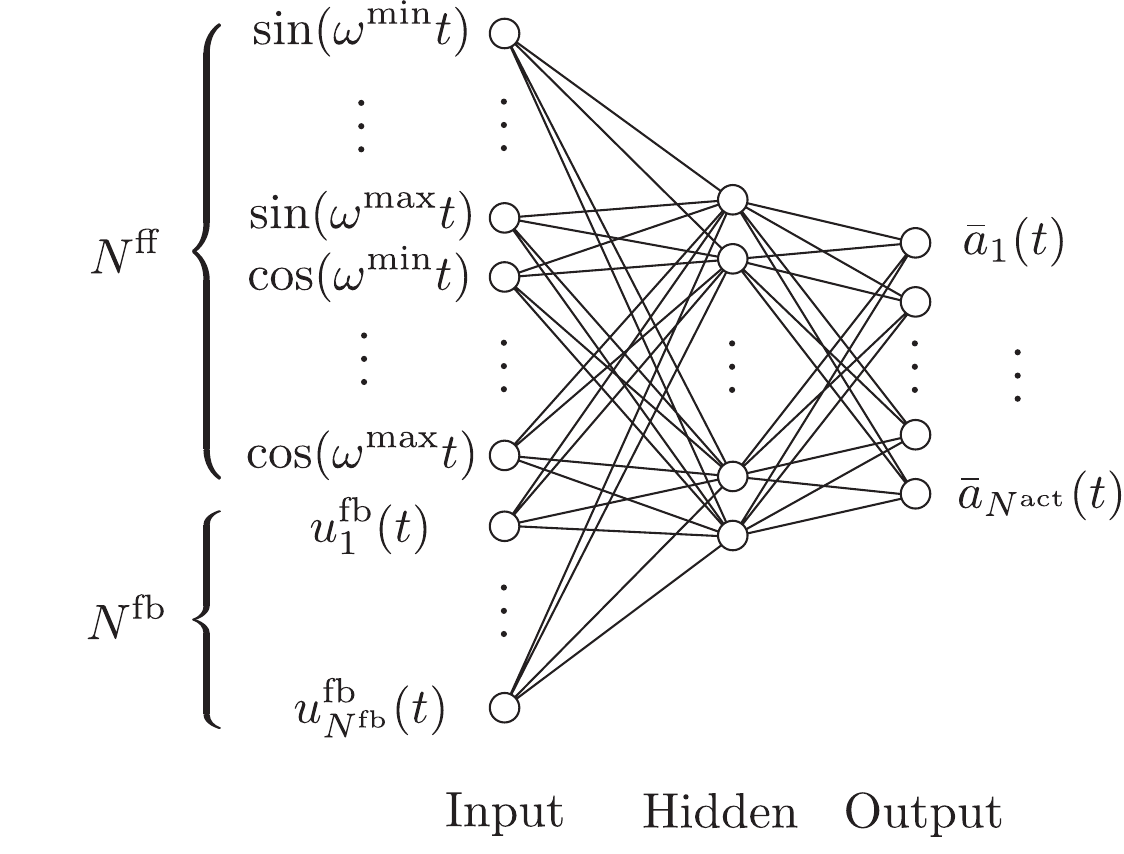}
    \caption{Neural network for the feedback controller. The input signals consist of feedforward and feedback signals where feedforward signals are given as sinusoidal and cosinusoidal functions, and the feedback signals are the terrain height surrounding the soft robots.}
    \label{fig:neural_network}
\end{figure}

\begin{figure}[t]
    \centering
    \includegraphics[width=0.5\textwidth]{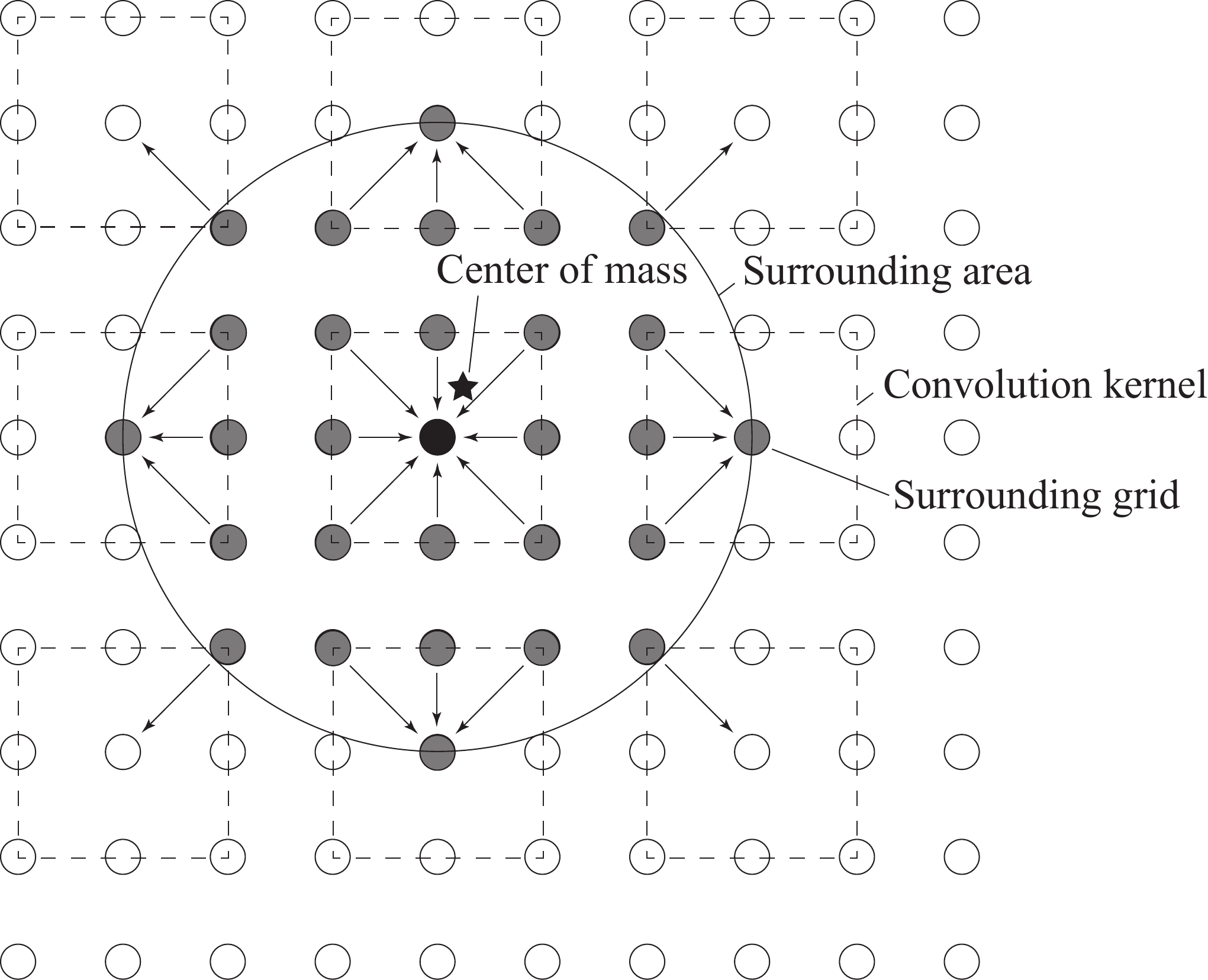}
    \caption{Conceptual diagram of convolution of terrain features. The star sign represents the center of mass of the soft robot and the black circle represents the grid node corresponding to the center of mass. A large circle defines the surrounding area and the terrain heights of the nodes in the circle (i.e., gray nodes whose indices form $\mathcal{B}^\mathrm{sur}(t)$) are used as feedback signals. Dashed boxes represent $3 \times 3$ neighboring nodes whose terrain heights are convoluted into the center node of each box, whose index forms $\tilde{\mathcal{B}}^\mathrm{sur}(t)$.}
    \label{fig:cnn}
\end{figure}

We now parametrize the actuation signal in the $j$-th channel, $\bar{a}_j$.
In this study, we use the terrain feature surrounding the soft body to design a feedback controller.
Since the terrain feature can be reflected in the boundary condition in Eq.~\eqref{eq:bnd_mpm}, it is reasonable to represent the terrain on the grid node.
Let $\bs{x}^\mathrm{grid}_m = (x^{\mathrm{grid}}_m, y^{\mathrm{grid}}_m, z^{\mathrm{grid}}_m)^\top$ denote the coordinate of the $m$-th grid node where the $y$-axis is taken along with the vertical upward direction.
Using the centroid of the material domain, $\bs{x}^\mathrm{cen}(t)$, defined as 
\begin{align}
    \bs{x}^\mathrm{cen}(t) := \frac{\sum_{i=1}^{N^\mathrm{par}} \gamma_i \bs{x}_i(t) }{\sum_{i=1}^{N^\mathrm{par}} \gamma_i}.
\end{align}
We can express the index set corresponding to the terrain surrounding the soft body, as follows:
\begin{align}
    &\mathcal{B}^\mathrm{sur}(t) := \left\{ m ~|~ m \in \mathcal{B}, ~ \| \texttt{proj}(\bs{x}^\mathrm{grid}_m) - \texttt{proj}(\bs{x}^\mathrm{cen}(t)) \| \leq \frac{W}{2} \right\},
\end{align}
where $W$ is the diameter to define the surrounding area around the centroid of the soft robot, and \texttt{proj} is the projector of positions onto the plane of $y=0$ where $y$-axis be taken along with the vertical upward direction.
To define a feedback controller based on this terrain feature, we use the neural network with one hidden layer.
Let $\bs{w}^{(1)}:=[w^{(1)}_{11}, w^{(1)}_{12}, \cdots, w^{(1)}_{N^\mathrm{hidden} N^\mathrm{input}}]$ and $\bs{w}^{(2)}:=[w^{(2)}_{11}, w^{(2)}_{12}, \cdots, w^{(1)}_{N^\mathrm{act} N^\mathrm{hidden}}]$ be the weights of the first and second layers, respectively.
Let $\bs{b}^{(1)}:=[b^{(1)}_{1}, \cdots, b^{(1)}_{N^\mathrm{hidden}}]$ and $\bs{b}^{(2)}:=[b^{(2)}_{1}, \cdots, b^{(2)}_{N^\mathrm{act}}]$ be the bias of the first and second layers, respectively.
The neural network that outputs the actuation signal $\bar{a}_{j}$ from the input $u_{l}$ through the hidden state $\tilde{u}_{k}$ is modeled, as follows:
\begin{align}
    \tilde{u}_{k}(t) &= \tanh \left( \sum_{l=1}^{N^\mathrm{input}} w^{(1)}_{kl} u_{l}(t) + b^{(1)}_{k} \right),  \\
    \bar{a}_{j}(t) &= c^\mathrm{act} \tanh \left( \sum_{k=1}^{N^\mathrm{hidden}} w^{(2)}_{jk} \tilde{u}_{k}(t) + b^{(2)}_{j} \right),
\end{align}
for $1 \leq l \leq N^\text{input},~ 1 \leq k \leq N^\text{hidden},~ 1 \leq j \leq N^\text{act}$ where 
$c^\mathrm{act}$ is a scaling factor of the actuation signal, $N^\text{input}$ and $N^\text{hidden}$ are the numbers of input and hidden nodes, respectively.
In addition to the feedback signals, we also use the feedforward signals as the input of the neural network.
Figure~\ref{fig:neural_network} illustrates the neural network used for the feedback controller.
Let $u^\text{ff}_l$ and $u^\text{fb}_l$ respectively denote the feedforward and feedback signals.
The input of the neural network is given, as follows:
\begin{align} \label{eq:nn_input}
    u_l(t) = \begin{cases}
        u^\text{ff}_l(t) & \text{if } 1 \leq l \leq N^\text{ff} \\
        u^\text{fb}_{l - N^\text{ff}}(t) & \text{if } N^\text{ff} < l \leq N^\text{ff} + N^\text{fb},
    \end{cases}
\end{align}
where $N^\text{ff}$ and $N^\text{fb}$ are the number of feedforward and feedback signals, respectively.
The feedforward signals consist of sinusoidal and cosinusoidal functions as
\begin{align}
    u^\text{ff}_l(t) = \begin{cases}
        \sin (\omega_{l} t) & \text{if } 1 \leq l \leq \lfloor\frac{N^\text{ff}}{2} \rfloor \\
        \cos (\omega_{l - \lfloor N^\text{ff}/2 \rfloor} t) & \text{if } \lfloor \frac{N^\text{ff}}{2} \rfloor < l \leq N^\text{ff},
    \end{cases}
\end{align}
where 
\begin{align}
    \omega_l := (\omega^\text{max} - \omega^\text{min}) \frac{l-1}{ \lceil N^\text{ff}/2 \rceil - 1} + \omega^\text{min}
\end{align}
with the maximum and minimum angular frequencies, denoted by $\omega^\text{max}$ and $\omega^\text{min}$, respectively.
The feedback signal is defined using the terrain feature surrounding the soft body, $\mathcal{B}^\mathrm{sur} (t)$, as follows:
\begin{align} \label{eq:fb_signal}
    u^\text{fb}_l (t) = \tanh \left( c^\mathrm{fb} \left( \boldsymbol{x}^{\text{cen}}(t) \cdot \boldsymbol{e}_y - y^\mathrm{grid}_{(\mathcal{B}^\mathrm{sur}(t))_l } - y^\mathrm{offset} \right) \right),
\end{align}
where $\bs{e}_y$ is the basis vector along with the $y$-axis, $c^\mathrm{fb}$ is a scaling factor, $y^\mathrm{offset}$ is the offset and $(\mathcal{B}^\mathrm{sur}(t))_l$ is the $l$-th element of $\mathcal{B}^\mathrm{sur}(t)$.
By this definition, the number of feedback signals coincides the number of elements of $\mathcal{B}^\mathrm{sur}(t)$, that is, $N^\mathrm{fb} = | \mathcal{B}^\mathrm{sur}(t) |$.
However, especially for three-dimensional problems, the number of feedback signals tends to be so large depending on the number of nodes in the grid.
Hence, for three-dimensional problems, we select representative nodes at every three nodes along the $x$ and $z$-axes from the index set $\mathcal{B}^\mathrm{sur}$, forming a new index set $\tilde{\mathcal{B}}^\mathrm{sur}$.
Then, we take the convolution of $3 \times 3$ nodes to reduce the number of signals, as follows:
\begin{align}
    \text{For } m \in \tilde{\mathcal{B}}^\mathrm{sur}(t), \quad \tilde{y}_{m}^\mathrm{grid} = \frac{1}{| \mathcal{I}_m^\mathrm{neigh} \cap \mathcal{B}^\mathrm{sur}(t) |} \sum_{m' \in \mathcal{I}_m^\mathrm{neigh} \cap \mathcal{B}^\mathrm{sur}(t) } y_{m'}^\mathrm{grid},
\end{align}
where $\mathcal{I}_m^\mathrm{neigh}$ represents the index set of $3 \times 3$ nodes neighboring the $m$-th node.
The feedback signal is given by replacing $y^\mathrm{grid}_{(\mathcal{B}^\mathrm{sur}(t))_l}$ in Eq.~\eqref{eq:fb_signal} with $\tilde{y}^\mathrm{grid}_{(\tilde{\mathcal{B}}^\mathrm{sur}(t))_l }$, as follows:
\begin{align}
    u^\text{fb}_l (t) = \tanh \left( c^\mathrm{fb} \left( \boldsymbol{x}^{\text{cen}}(t) \cdot \boldsymbol{e}_y - \tilde{y}^\mathrm{grid}_{(\tilde{\mathcal{B}}^\mathrm{sur}(t))_l } - y^\mathrm{offset} \right) \right).
\end{align}
Figure~\ref{fig:cnn} illustrates the concept of the convolution of terrain fearures.
This is a kind of the convolutional neural network (CNN) and we can also use more sophisticated CNNs for this purpose.
Design variables for the neural network is summarized as $\bs{w}:=[\bs{w}^{(1)}, \bs{b}^{(1)}, \bs{w}^{(2)}, \bs{b}^{(2)}]$.

\subsection{Objective function} \label{sec:objective_function}
Next, we formulate the objective function.
Let $L$ denote the target travel distance during a time duration $T$.
The objective function $\mathcal{F}$ for a soft body to achieve the target travel distance keeping the posture is formulated, as follows:
\begin{align} \label{eq:objective_orig}
    \mathcal{F}(\bs{\phi}, \bs{\psi}, \bs{w}) := \mathrm{min} \left( 1 - \frac{\int_{0}^{T} \bs{v}^\mathrm{cen}(t) \cdot \bs{e}_x \mathrm{d}t}{TL}, 0 \right)^2 + \frac{w_\theta}{T} \int_{0}^{T} \left\| \frac{\bs{\theta}(t)}{\bar{\theta}} \right\|^2 \mathrm{d}t,
\end{align}
where $w_\theta$ is a weighting coefficient and $\bs{v}^\mathrm{cen}$ is the velocity of the centroid defined as
\begin{align}
    \bs{v}^\mathrm{cen}(t) := \frac{\sum_{i=1}^{N^\mathrm{par}} \gamma_i \bs{v}_i(t) }{\sum_{i=1}^{N^\mathrm{par}} \gamma_i},
\end{align}
where $\bs{\theta}$ is the rotation vector and $\bar{\theta}$ is a normalization parameter.
Note that the velocity $\bs{v}_i$ and the rotation vector $\bs{\theta}$ implicitly depend on the design variables $\bs{\phi}$, $\bs{\psi}$ and $\bs{w}$ through the forward problem.
In the following, we explain how to calculate the rotation vector $\bs{\theta}$. 
Let $\bs{J}$ denote the inertia matrix around the centroid.
The angular velocity vector $\bs{\omega}$ is then given as
\begin{align}
    \bs{\omega}(t) = \bs{J}^{-1}(t) \sum_{i=1}^{N^\mathrm{par}} \rho_i (\bs{x}_i(t) - \bs{x}^\mathrm{cen}(t)) \times (\bs{v}_i(t) - \bs{v}^\mathrm{cen}(t)).
\end{align}
The unit quaternion $\bs{q}(t)$ describing the posture of the soft body is given as
\begin{align}
    \bs{q}(t) = \bs{q}(0) + \frac{1}{2} \int_0^t \bs{\omega}(t') \bs{q}(t') \mathrm{d} t',
\end{align}
where $\bs{\omega}(t') \bs{q}(t')$ is the product of quaternions.
Then, the rotation vector is calculated by the quaternion, as follows:
\begin{align}
    \bs{\theta}(t) = \frac{2\tan^{-1} \left( \frac{\sqrt{q_1(t)^2 + q_2(t)^2 + q_3(t)^2}}{q_0(t)} \right) }{\sqrt{q_1(t)^2 + q_2(t)^2 + q_3(t)^2}} \left[ q_1(t), q_2(t), q_3(t) \right]^\top,
\end{align}
where $q_0, q_1, q_2, q_3$ are components of the quaternion $\bs{q}$.
By minimizing the objective function $\mathcal{F}$, we obtain the optimized structure, actuator layout, and neural network-based controller of a soft body for a prescribed terrain given by $y_m^\mathrm{grid}$.

Since we would like to design a soft body which can travel under terrain uncertainty, we prepare dataset of terrains and \textit{train} the soft body using the dataset based on the machine learning approach.
Let $\mathcal{F}^{(n)}$ be the objective function for the $n$-th terrain data.
Then, the optimization problem to train the soft body by $N^\mathrm{data}$ training data is formulated, as follows:
\begin{alignat}{2}
    &\underset{\bs{\phi}, \bs{\psi}, \bs{w}}{\text{minimize}} \quad \quad && \frac{1}{N^\mathrm{data}} \sum_{n=1}^{N^\mathrm{data}} \mathcal{F}^{(n)}(\bs{\phi}, \bs{\psi}, \bs{w}) + \frac{1}{2} \alpha \left\| \bs{w} \right\|^2 \label{eq:objective} \\
    &\text{subject to:} && \mathcal{C}^\mathrm{to}(\bs{\phi}) \leq \bar{C}^\mathrm{to} \label{eq:const_to} \\
    & &&\mathcal{C}^\mathrm{lay}(\bs{\psi}) \leq \bar{C}^\mathrm{lay}, \label{eq:const_lay}
\end{alignat}
where $\alpha$ is a coefficient for L2 regularization of the neural network and $\mathcal{C}^\mathrm{to}$ and $\mathcal{C}^\mathrm{lay}$ are the constraint functions defined as
\begin{align}
    \mathcal{C}^\mathrm{to}(\bs{\phi}) &:= \frac{4}{N^\mathrm{par}} \sum_{i=1}^{N^\mathrm{par}} \gamma_i (\phi_i) (1 - \gamma_i (\phi_i)) \\
    \mathcal{C}^\mathrm{lay}(\bs{\psi}) &:= \frac{N^\mathrm{act} + 1}{N^\mathrm{par} N^\mathrm{act}} \sum_{i=1}^{N^\mathrm{par}} \left( 1 - \sum_{j=1}^{N^\mathrm{act}+1} \xi_{ij}(\psi_{ij})^2 \right),
\end{align}
for ensuring the binarization of the material density $\gamma_i$ and relaxed two-index variable $\xi_{ij}$ by setting the upper bounds $\bar{C}^\mathrm{to}$ and $\bar{C}^\mathrm{lay}$ to a small value.
Here, we used the notation $\gamma_i (\phi_i)$ and $\xi_{ij}(\psi_{ij})$ to explicitly show that they depend on design variables $\phi_i$ and $\psi_{ij}$ through the particle-wise filtering and Heaviside projection/softmax projection, respectively.
The function $\mathcal{C}^\mathrm{to}(\bs{\phi}) \in [0, 1]$ has the minimum value of zero when $\gamma_i$ takes either zero or one for all particles while it has the maximum value of one when $\gamma_i = 0.5$ for all particles.
The function $\mathcal{C}^\mathrm{lay}(\bs{\psi}) \in [0, 1]$ has the minimum value of zero when $\xi_{ij}=1$ for a value of $j$ and $\xi_{ij'}=0$ for all $j' \neq j$ for all particles while it has the maximum value of one when $\xi_{ij} = 1 / (N^\mathrm{act} + 1)$ for all $j$ and all particles.
That is, $\mathcal{C}^\mathrm{to}$ and $\mathcal{C}^\mathrm{lay}$ represent the ratio of intermediate values of $\gamma_i$ and $\xi_{ij}$, respectively.

\subsection{Dataset}
For training a soft body, we prepare terrain dataset by randomly sampling terrains from Gaussian process.
Let a box $[0, L_x] \times [0, L_y] \times [0, L_z]$ denote the simulation environment with the coordinate $\bs{x}=[x, y, z]$.
We take $y$-axis along with the vertical upward direction as in Sec.~\ref{sec:controller}.
Let $\mathcal{K}$ be the Gaussian kernel defined as
\begin{align}
    \mathcal{K}(\bs{x}, \bs{x}') := \exp \left( - \frac{ (x - x')^2 + (z - z')^2 }{2\varrho^2} \right),
\end{align}
where $\varrho$ is a hyperparameter determining the radius of the kernel.
We then sample a function describing the height of terrains from the Gaussian process $\mathcal{GP}(\bar{h}, \mathcal{K})$ with the constant mean function $\bar{h}$ and the kernel function $\mathcal{K}$, as follows:
\begin{align}
    h(x, z) \sim \mathcal{GP}(\bar{h}, \varsigma \mathcal{K}),
\end{align}
where $\varsigma$ is a hyperparameter for scaling the height difference in a terrain.
The function $h(x, z)$ gives the height of a terrain at a horizontal coordinate $(x, z)$.
For performing the forward simulation using each terrain data, we first have to put a soft body on an initial position, which should be fixed for all terrain data to eliminate the influence of initial positions.
However, since the function $h(x, z)$ gives the random height all over the environment, it is difficult to determine an initial position available for all data.
Thus, we fix the height of terrains on the area satisfying $x \leq \bar{x}$ to $\bar{h}$ and set the initial position on the area.
To make the terrain continuously connect to this fixed area, we modify the function $h(x, z)$, as follows:
\begin{align}
    \tilde{h}(x, z) = h(x, z) - \frac{h(L_x, z) - h(\bar{x}, z)}{L_x - \bar{x}}(x - \bar{x}) - h(\bar{x}, z) + \bar{h}.
\end{align}
The modified function $\tilde{h}(x, z)$ satisfies $\tilde{h}(\bar{x}, z) = \tilde{h}(L_x, z) = \bar{h}$.
We eventually obtain an index set describing grid nodes on the ground for the boundary condition in Eq.~\eqref{eq:bnd_mpm}, as follows:
\begin{align}
    \mathcal{B} = \left\{m ~|~ y^\mathrm{grid}_m = \left\lfloor \frac{\tilde{h}(x^\mathrm{grid}_m, z^\mathrm{grid}_m)}{\Delta_y} \right\rfloor \Delta_y \right\},
\end{align}
where $\Delta_y$ is the interval between grid nodes along with the $y$-axis.
The floor function maps the continuous height onto the discrete grid.
We also prepare test dataset as well as training dataset.

\begin{algorithm}[t]
    \SetKwInOut{Input}{Input}
    \SetKwInOut{Output}{Output}

    \Input{Initial design variables $\bs{\phi}^0$, $\bs{\psi}^0$ and $\bs{w}^0$, penalty coefficient $\bs{\tau}^0$.}
    \Output{Optimized structure $[ \gamma_1, \dots, \gamma_{\mathrm{N}^\mathrm{par}} ]$, actuator layout $[\xi_{11}, \cdots, \xi_{N^\text{par} N^\text{act}+1}]$, neural network parameters $\bs{w}$}
    \BlankLine

    Divide $N^\mathrm{data}$ into $N^\mathrm{batch}$ batches $\{\mathcal{I}^\mathrm{batch}_n \}_{n=1}^{N^\mathrm{batch}}$ randomly. \\
    Set $s \leftarrow 0$, $n \leftarrow 1$, $\bs{\kappa} \leftarrow \bs{0}$, $\bs{\tau} \leftarrow \bs{\tau}^0$. \\
    Set \textit{flag} $\leftarrow$ True. \\
    
    \While{flag}{
        Compute $\mathcal{L}_n (\bs{\phi}^s, \bs{\psi}^s, \bs{w}^s, \bs{\kappa}, \bs{\tau})$ and its gradients. \\
        Set $\mathcal{L}^s \leftarrow \mathcal{L}_n$. \\
        
        \uIf{$\left| \sum_{s'=s-N^\mathrm{batch} + 1}^{s} \mathcal{L}^{s'} - \sum_{s'=s-2N^\mathrm{batch} + 1}^{s-N^\mathrm{batch}} \mathcal{L}^{s'} \right| < \epsilon$}{
            \uIf{$\mathcal{C}^\mathrm{to} \leq \bar{C}^\mathrm{to}$ and $\mathcal{C}^\mathrm{lay} \leq \bar{C}^\mathrm{lay}$}{
                Set \textit{flag} $\leftarrow$ False. \\
                Return $\{ \gamma_i (\phi_i^s) \}_i$, $\{ \xi_{ij}(\psi_i^s) \}_{ij}$, $\bs{w}^s$.
            }\uElse{
                Update Lagrange multiplier $\bs{\kappa}$ and penalty coefficient $\bs{\tau}$.
            }
        }
        
        Update design variables to $\bs{\phi}^{s+1}$, $\bs{\psi}^{s+1}$ and $\bs{w}^{s+1}$ by Adam. \\
        
        \uIf{$n=N^\mathrm{batch}$}{
            Divide $N^\mathrm{data}$ into $N^\mathrm{batch}$ batches $\{\mathcal{I}^\mathrm{batch}_n \}_{n=1}^{N^\mathrm{batch}}$ randomly. \\
            Set $n \leftarrow 0$.
        }
        Set $s \leftarrow s + 1$, $n \leftarrow n + 1$. \\
    }

 \caption{Optimization algorithm}\label{alg}
\end{algorithm}

\subsection{Optimization algorithm}

To solve the optimization problem formulated in Sec.~\ref{sec:objective_function}, we use the augmented Lagrangian method~\cite{bertsekas2014constrained}, which converts a constrained problem to an unconstrained problem.
As an optimizer, we use the Adam~\cite{kingma2015adam}, a kind of stochastic gradient methods, with mini-batch learning.
That is, we pick up partial $N^\mathrm{batch}$ data at an iteration to evaluate the objective function and its gradients and to update the design variables.

Let $\mathcal{I}^\mathrm{batch}_n$ denote the index set of training data in the $n$-th batch with the size $N^\mathrm{batch}$.
We assume that all data $N^\mathrm{data}$ can be divided into $N^\mathrm{batch}$ batches without residue satisfying $\bigcup_{n=1}^{N^\mathrm{batch}} \mathcal{I}^\mathrm{batch}_n = \{1, 2, \cdots, N^\mathrm{data} \}$.
The augmented Lagrangian $\mathcal{L}$ for the $n$-th batch is formulated, as follows:
\begin{align}
    &\mathcal{L}_n(\bs{\phi}, \bs{\psi}, \bs{w}, \bs{\kappa}, \bs{\tau}) \nonumber \\
    &:= \frac{1}{|\mathcal{I}^\mathrm{batch}_n | }\sum_{n' \in \mathcal{I}^\mathrm{batch}_n} \mathcal{F}^{(n')} + \frac{1}{2} \alpha \left\| \bs{w} \right\|^2 \nonumber \\
    & \quad - \kappa^\mathrm{to} \max \left( \mathcal{C}^\mathrm{to}(\bs{\phi}) - \bar{C}^\mathrm{to}, 0 \right) + \frac{1}{2} \tau^\mathrm{to}\max \left( \mathcal{C}^\mathrm{to}(\bs{\phi}) - \bar{C}^\mathrm{to}, 0 \right)^2  \nonumber \\
    & \quad - \kappa^\mathrm{lay} \max \left( \mathcal{C}^\mathrm{lay}(\bs{\psi}) - \bar{C}^\mathrm{lay}, 0 \right) + \frac{1}{2} \tau^\mathrm{lay}\max \left( \mathcal{C}^\mathrm{lay}(\bs{\psi}) - \bar{C}^\mathrm{lay}, 0 \right)^2,
\end{align}
where $\bs{\kappa} := [\kappa^\mathrm{to}, \kappa^\mathrm{lay}]$ is the Lagrange multiplier and $\bs{\tau} := [\tau^\mathrm{to}, \tau^\mathrm{lay}]$ is the penalty coefficient.
Since the average of the augmented Lagrangians for all batches corresponds to the augmented Lagrangian for all data, that is,
\begin{align}
    &\frac{1}{N^\mathrm{batch}} \sum_{n=1}^{N^\mathrm{batch}} \mathcal{L}_n \nonumber \\
    &= \frac{1}{N^\mathrm{data} }\sum_{n'=1}^{N^\mathrm{data}} \mathcal{F}^{(n')} + \frac{1}{2} \alpha \left\| \bs{w} \right\|^2 \nonumber \\
    & \quad - \kappa^\mathrm{to} \max \left( \mathcal{C}^\mathrm{to}(\bs{\phi}) - \bar{C}^\mathrm{to}, 0 \right) + \frac{1}{2} \tau^\mathrm{to}\max \left( \mathcal{C}^\mathrm{to}(\bs{\phi}) - \bar{C}^\mathrm{to}, 0 \right)^2  \nonumber \\
    & \quad - \kappa^\mathrm{lay} \max \left( \mathcal{C}^\mathrm{lay}(\bs{\psi}) - \bar{C}^\mathrm{lay}, 0 \right) + \frac{1}{2} \tau^\mathrm{lay}\max \left( \mathcal{C}^\mathrm{lay}(\bs{\psi}) - \bar{C}^\mathrm{lay}, 0 \right)^2,
\end{align}
the stochastic gradient approach works to solve the problems in Eqs.~\eqref{eq:objective}--\eqref{eq:const_lay} by the augmented Lagrangian method and mini-batch learning.
We describe the optimization algorithm in Algorithm~\ref{alg}.
We implemented this algorithm using Python and Taichi~\cite{hu2019taichi, hu2019difftaichi}, an open-source programming language which supports parallel computing on GPU and provides an automatic differentiation (AD) scheme.
We used AD to compute the gradient in line 5 in the algorithm.

\section{Numerical experiments} \label{sec:experiments}
We now present two numerical experiments, where we design a soft body that moves toward $x$-direction, which we named \textit{Walker}, in two- and three-dimensions.
We executed our algorithms on the GPU (NVIDIA A100 40GB) for all examples. 

\subsection{Hyperparameter settings}
We first describe hyperparameters using the same values across all numerical experiments. 
Hyperparameters set on specific problems are provided in the respective examples.

\subsubsection{Material property}
The soft body consisted of a soft silicone rubber with the density $\bar{\rho}=1000~\mathrm{kg/m^3}$, the Young's modulus $E=0.1~\mathrm{MPa}$ and the Poisson ratio $\nu = 0.4$, which have the relationship with the Lam\'{e} constants that
\begin{align}
    \bar{\lambda} &= \frac{E \nu}{(1 + \nu) (1 - 2\nu)} \\
    \bar{\mu} &= \frac{E}{2 (1 + \nu)}.
\end{align}
We set the small constant $\varepsilon=10^{-5}$ for numerical stability.

\subsubsection{Forward problem settings}
The computational domain representing simulation environment consisted of a square and a cube in 2D and 3D problems, respectively, with a length of $L_x = L_y = L_z = 1~\mathrm{m}$.
We discretized the domain by a grid with equal spacing $1.25 \times 10^{-2}~\mathrm{m}$.
We placed particles for the MPM at $6.25 \times 10^{-3} \mathrm{m}$ intervals, which is half the length of the grid spacing.
We used $\eta = 1 \times 10^{2}~\mathrm{Pa \cdot s}$ for the damping parameter to mitigate high-frequency oscillations.
The simulation time was set to $T = 1~\mathrm{s}$ which we divided with a time step of $\Delta_t = 1 \times 10^{-4}~\mathrm{s}$.
The gravity acceleration was set to $\bs{g}=[0, -9.8~\mathrm{m/s^2}, 0]^\top$.

\subsubsection{Hyperparameters for the neural network}
We initialized neural network parameters for the controller by sampling from the Gaussian distribution $\mathcal{N}$, based on the Xavier initialization scheme~\cite{glorot2010understanding}, as follows: 
\begin{align}
    & w_{kl}^{(1)} \sim \mathcal{N} \left(0, \sqrt{\frac{2}{N^\mathrm{hidden} + N^\mathrm{input}}} \right) \\
    & b_k^{(1)} = 0 \\
    & w_{jk}^{(2)} \sim \mathcal{N} \left(0, \sqrt{\frac{2}{N^\mathrm{act} + 1 + N^\mathrm{hidden}}} \right) \\
    & b_j^{(2)} = 0,
\end{align}
which serves for the outputs and gradients of neural networks to flow effectively in both forward and back propagation.
We set $\alpha=0.01$, the regularization coefficient for the neural network.
We prepared $N^\mathrm{ff}=56$ feedforward signals with the maximum and minimum angular frequencies configured to $\omega^\mathrm{max} = 8 \times 2\pi$ and $\omega^\mathrm{min} = 2 \times 2\pi$, respectively.
We configured the offset of feedback signals as $y^\mathrm{offset} = \bs{x}^\mathrm{cen}(0) \cdot \bs{e}_y - \bar{h}$, which calibrated the feedback signals at an initial time to zero.
The number of hidden nodes was determined by $N^\mathrm{hidden} = \lfloor (N^\mathrm{input} + N^\mathrm{act} + 1) / 2 \rfloor$.
The actuation signal is scaled by $c^\mathrm{act} = 2 \times 10^4~\mathrm{Pa}$.

\subsubsection{Hyperparameters for optimization}
We set initial design variables for the structure and actuator layout as $\bs{\phi}=\bs{0}$ and $\bs{\psi}=\bs{0}$.
We also set $R=9.375 \times 10^{-3} \mathrm{m}$, which is one and a half times than particle intervals, $p=2$, $\beta^\mathrm{to}=4$ and $\beta^\mathrm{lay}=4$ for the filtering and projection of design variables to fictitious density $\{\gamma_i\}$ and relaxed two-index variables $\{\xi_{ij}\}$.
For the objective function, we set $\bar{\theta} = \pi/2$, which is the normalization parameter of the angle describing posture of the soft body.
As for the augmented Lagrangian method, we assigned $\bs{\tau}^0 = [\tau^\mathrm{to}, \tau^\mathrm{lay}] = [0.001, 0.001]$ and these values were updated in line 12 of Algorithm~\ref{alg} based on the same strategy with our previous work~\cite{yuhn20234d}.
We set the learning rate of Adam to $0.02$.

\subsection{2D walker}
\begin{figure}[t]
    \centering
    \includegraphics[width=0.6\textwidth]{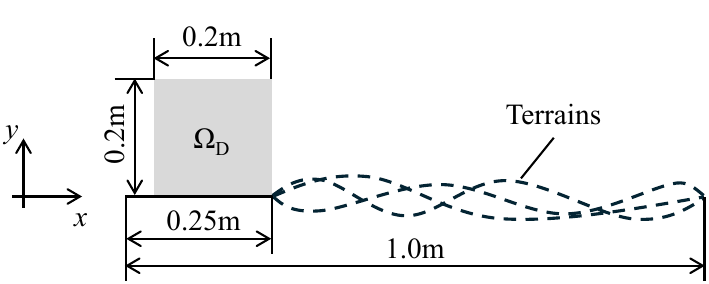}
    \caption{Problem setting for the 2D walker. The gray square represents the design domain of a soft robot and dashed line illustrates the schematic diagram of randomly generated terrains. The soft robot is optimized so that the gray square can travel toward the right direction.}
    \label{fig:walker_2d}
\end{figure}

We first designed a soft robot that moved toward $x$-direction in two-dimensions.
We configured the fixed design domain $\Omega_\mathrm{D}$ as a $0.2 \times 0.2 \mathrm{m}^2$ square on a flat ground as shown in Fig.~\ref{fig:walker_2d}.
We set the number of actuators as $N^\mathrm{act} = 4$.
For terrain dataset, we assigned $\varsigma=0.02~\mathrm{m}$ and $\varrho=0.2~\mathrm{m}$, respectively.
We empirically set $w_\theta=1$ for the weighting coefficient in the objective function \eqref{eq:objective_orig}.
We set $c^\mathrm{fb} = 1 / \varsigma = 50~\mathrm{m^{-1}}$, the parameter for scaling the feedback signals in Eq.~\eqref{eq:fb_signal}, which normalized the feedback signals independent of the terrain heights.

\subsubsection{Effect of size of surrounding area}
First, we examined the effect of the size of the surrounding area, $W$, to be fed into the feedback controller.
For the target travel distance, we set $L=0.7~\mathrm{m}$.
We prepared $32 (=N^\mathrm{data})$ training data and also prepared the same number of test data.

\begin{figure}[t]
    \centering
    \includegraphics[width=0.7\textwidth]{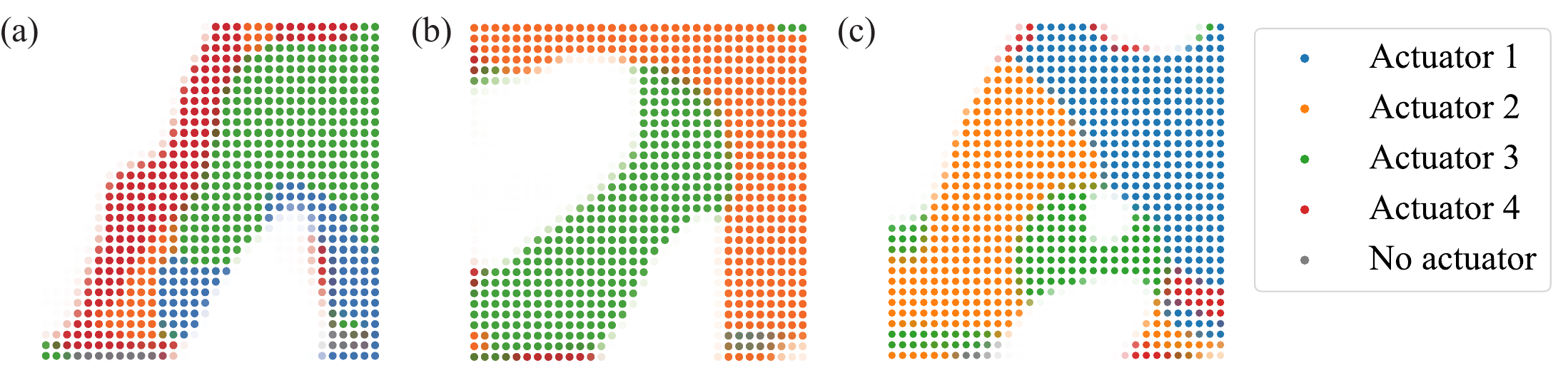}
    \caption{Optimized structures and actuator layouts for various settings of $W$. (a)~$W=0$, (b)~$W=0.3$ and (c)~$W=0.4$.}
    \label{fig:walker_2d_result_ex1}
\end{figure}

Figure~\ref{fig:walker_2d_result_ex1} illustrates the optimized structures and actuation layouts obtained for various settings of the surrounding area $W$.
The case with $W = 0$ means the soft robot employs no feedback controller and just uses feedforward signals $u_l^\mathrm{ff}$ in Eq.~\eqref{eq:nn_input}.
All optimized structures in Fig.~\ref{fig:walker_2d_result_ex1} have a common feature of two leg-like substructures while detailed parts differ.
Actuator layouts on the structures also share similarities; the left and right \textit{legs} are equipped with different actuators.
These features also appeared in the previous study~\cite{yuhn20234d} and seem to be essential to walk.

\begin{figure}[t]
    \centering
    \includegraphics[width=0.7\textwidth]{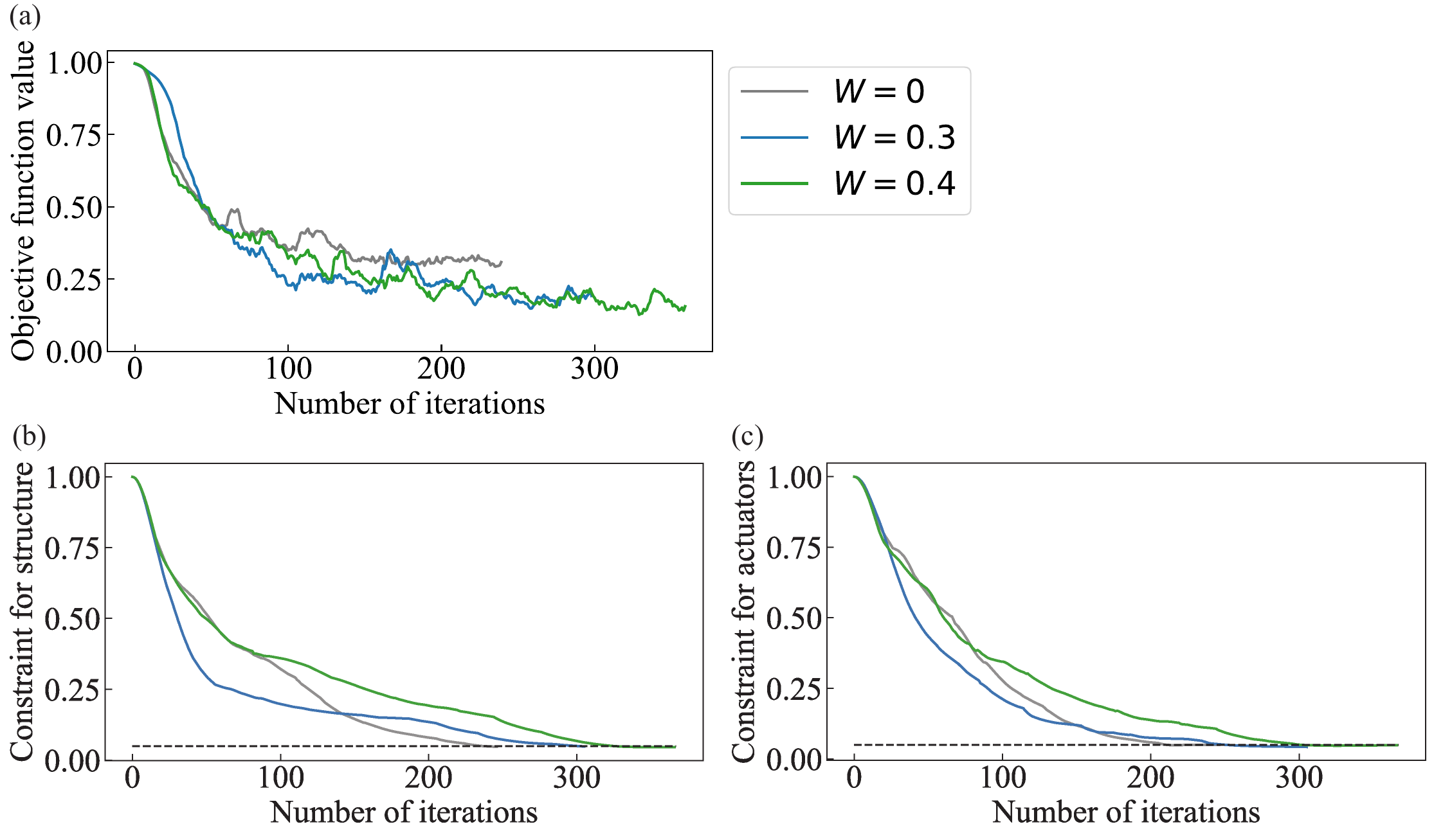}
    \caption{Trajectories of (a) objective function, (b) constraint function for binarization of structure and (c) constraint function for binarization of actuator layout during optimization with respect to the surrounding area width. The objective function values are plotted with a moving average taken every $N^\mathrm{data} / N^\mathrm{batch}(=8)$ iterations. Dashed lines in (b) and (c) represents the prescribed upper bounds of the constraint functions.}
    \label{fig:walker_2d_history_ex1}
\end{figure}

Figure~\ref{fig:walker_2d_history_ex1} shows the trajectories of the objective function $\mathcal{F}$ and constraint functions for binarization of material density and relaxed two-index variables, $\mathcal{C}^\mathrm{to}$ and $\mathcal{C}^\mathrm{lay}$, over the optimization iterations where the objective function was plotted with a moving average taken every $N^\mathrm{data} / N^\mathrm{batch}(=8)$ iteration.
For all settings of $W$ values, the objective function decreased from 1.0 and stayed around 0.25 with oscillation while constraint functions smoothly reached the upper bounds represented by the dashed lines to satisfy the constraints.
Oscillations in the objective function values stem from the stochastic gradient method where the design variables were updated by using different training data at every iteration.
No such oscillation was observed for the constraint functions due to the independence of the constraint function from the training dataset.
These figures show that our algorithm could decrease the objective function in a stochastic manner while dealing with the constraint functions for structural and actuator layout optimization.
The objective function values in the cases of $W=0.3$ and $0.4$ were lower than that of $W=0$ on average, which implies that the feedback controller contributes to soft robots walking on varying terrains more stably.

\begin{figure}[t]
    \centering
    \includegraphics[width=0.8\textwidth]{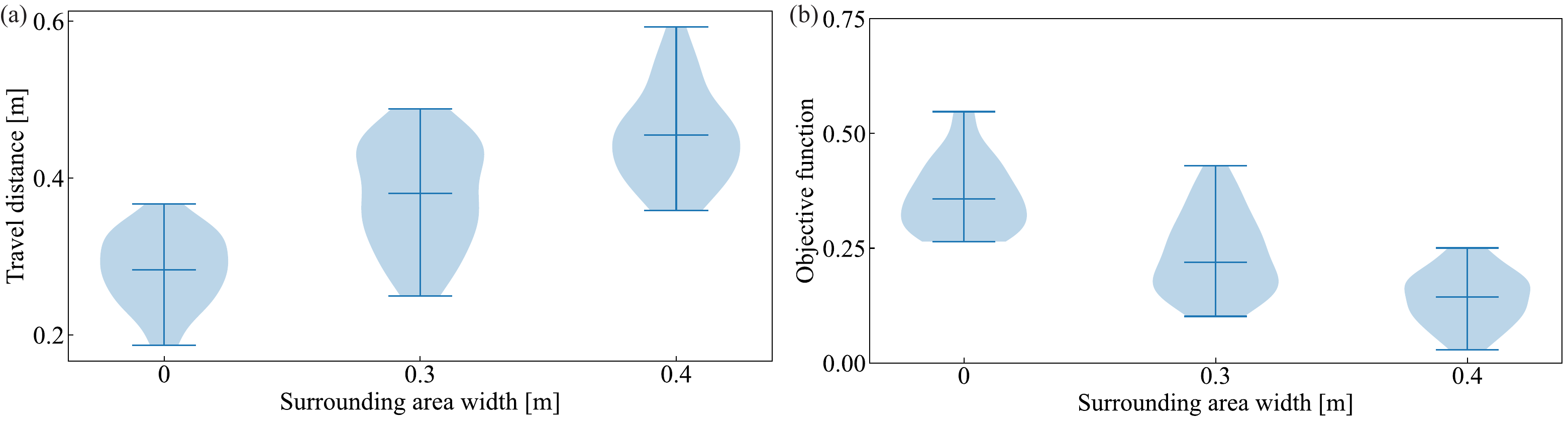}
    \caption{Violin plots illustrating (a) travel distances and (b) objective function values of the optimized soft robots for the test dataset with respect to the surrounding area width $W$.}
    \label{fig:walker_2d_violin_ex1}
\end{figure}

Next, we discussed the result using the test dataset which was not used in the training (optimization) procedure.
Figure~\ref{fig:walker_2d_violin_ex1} shows violin plots illustrating the travel distances and objective function values of the optimized soft robots in Fig.~\ref{fig:walker_2d_result_ex1}, evaluated for 32 test data.
Figure~\ref{fig:walker_2d_violin_ex1}(a) indicates that the travel distances for $W=0.3$ and $0.4$ were larger on average than that for $W=0$, which suggests that the feedback controller contributes to improving the travel distance.
In terms of the objective function, which includes the term for keeping the posture in addition to increasing the travel distance, the cases of $W=0.3$ and $0.4$ resulted in better performance on average than the case of $W=0$ as well.
The objective function values evaluated using the test dataset roughly range around 0.25, closely aligning with those obtained from the training dataset. 
This confirmation assures us that the training process did not result in overfitting.

\begin{figure}[t]
    \centering
    \includegraphics[width=0.8\textwidth]{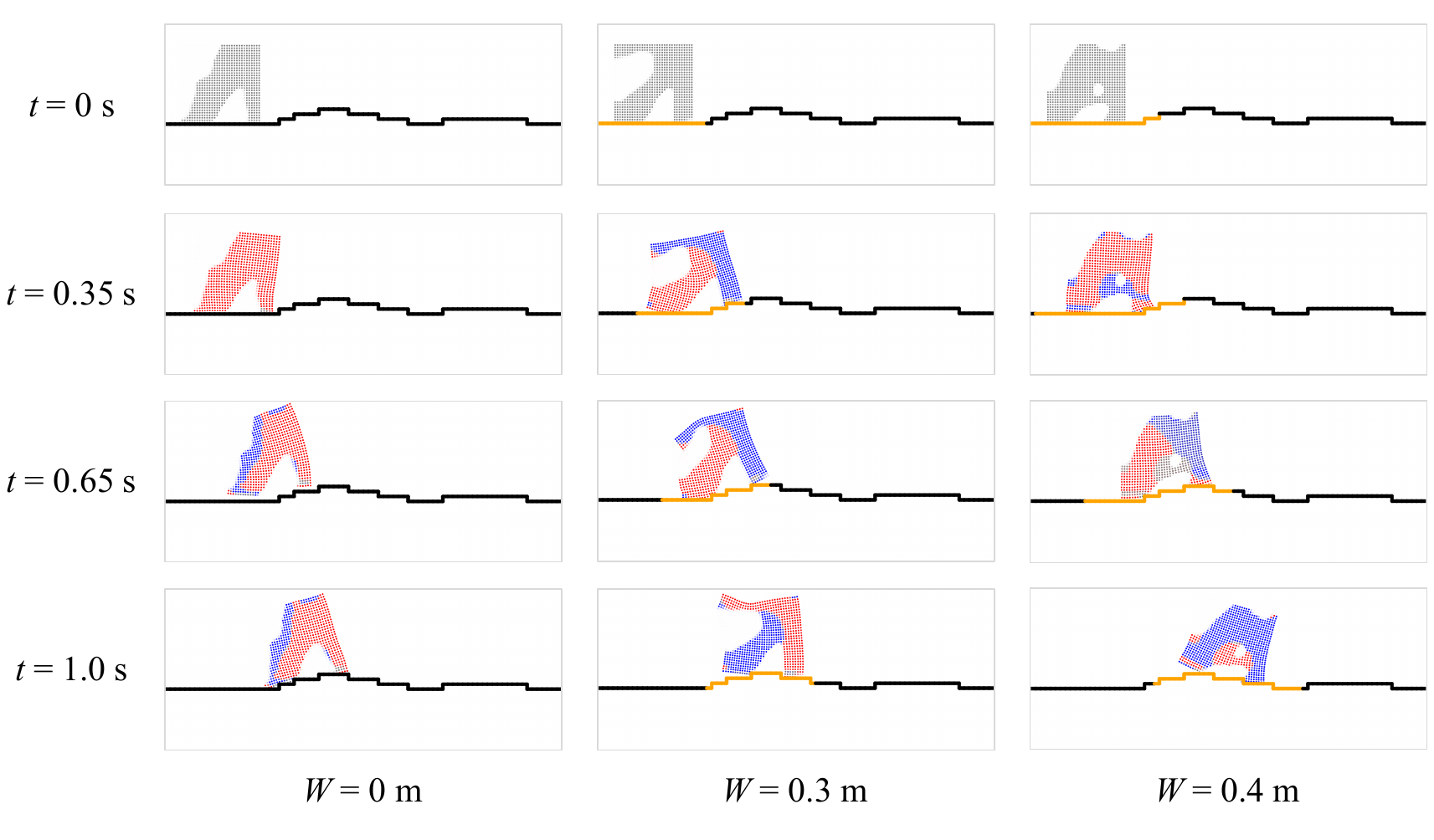}
    \caption{Dynamic behaviors of the three optimized soft robots for the test terrain on which the largest objective function was observed among the test dataset for the soft robot equipped with a feedforward controller. The red and blue represent the expansion and contraction by actuation, respectively. The black line represents the terrain where the orange segment is the range used as the input of the controller.}
    \label{fig:walker_2d_ex1_test23_wst}
\end{figure}

\begin{figure}[t]
    \centering
    \includegraphics[width=0.55\textwidth]{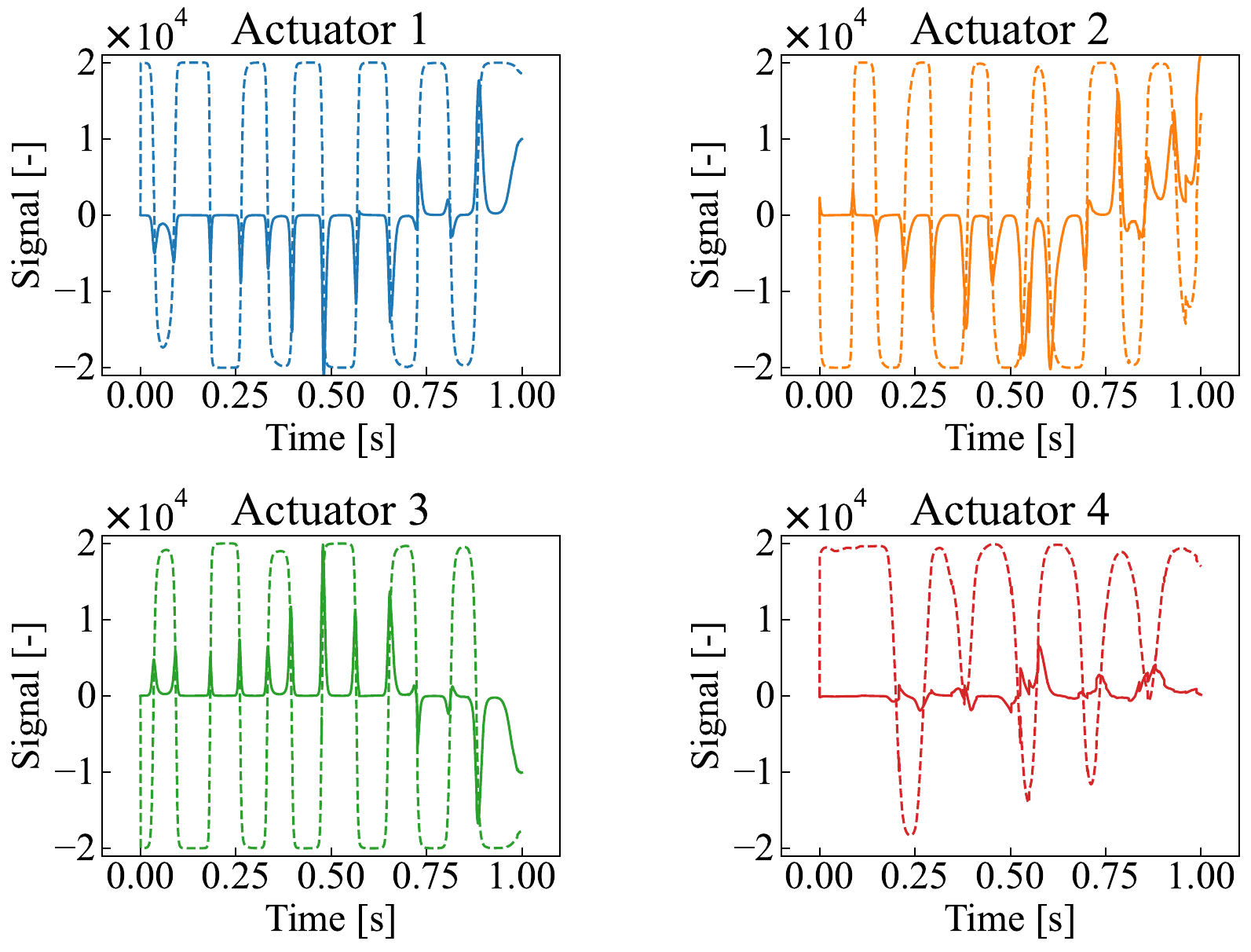}
    \caption{Signals of the respective actuators over the simulation duration. Each line represents the deviation of each actuation signal from its mean value of all test data. Each dashed line represents the actuation signal. Positive and negative actuation signals represent expansion and contraction, respectively.}
    \label{fig:walker_2d_ex1_act_test23_wst}
\end{figure}

\begin{figure}[t]
    \centering
    \includegraphics[width=0.8\textwidth]{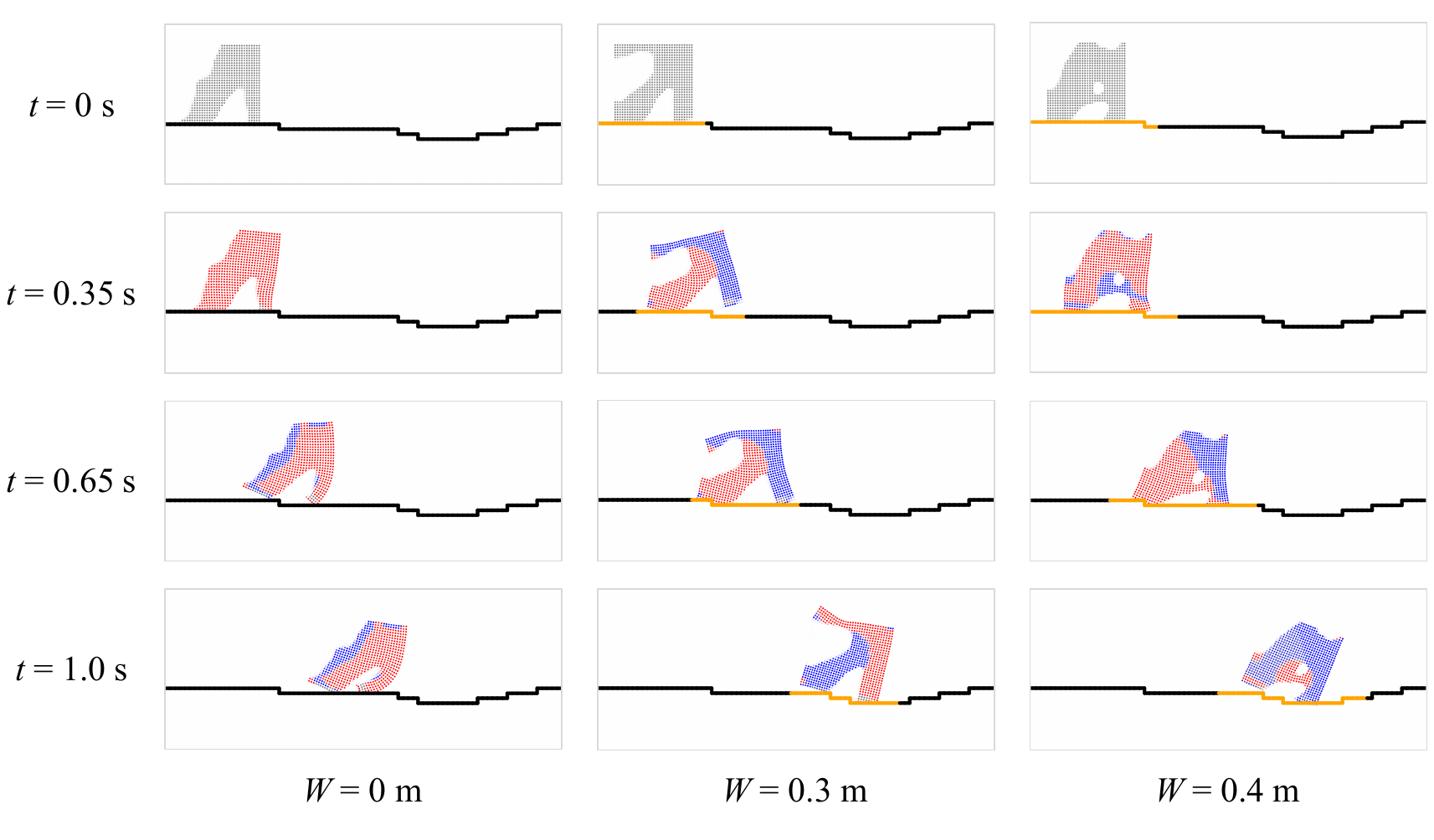}
    \caption{Dynamic behaviors of the three optimized soft robots for the test terrain on which the smallest objective function was observed among the test dataset for the soft robot equipped with a feedforward controller. The red and blue represent the expansion and contraction by actuation, respectively. The black line represents the terrain where the orange segment is the range used as the input of the controller.}
    \label{fig:walker_2d_ex1_test11_bst}
\end{figure}

\begin{figure}[t]
    \centering
    \includegraphics[width=0.55\textwidth]{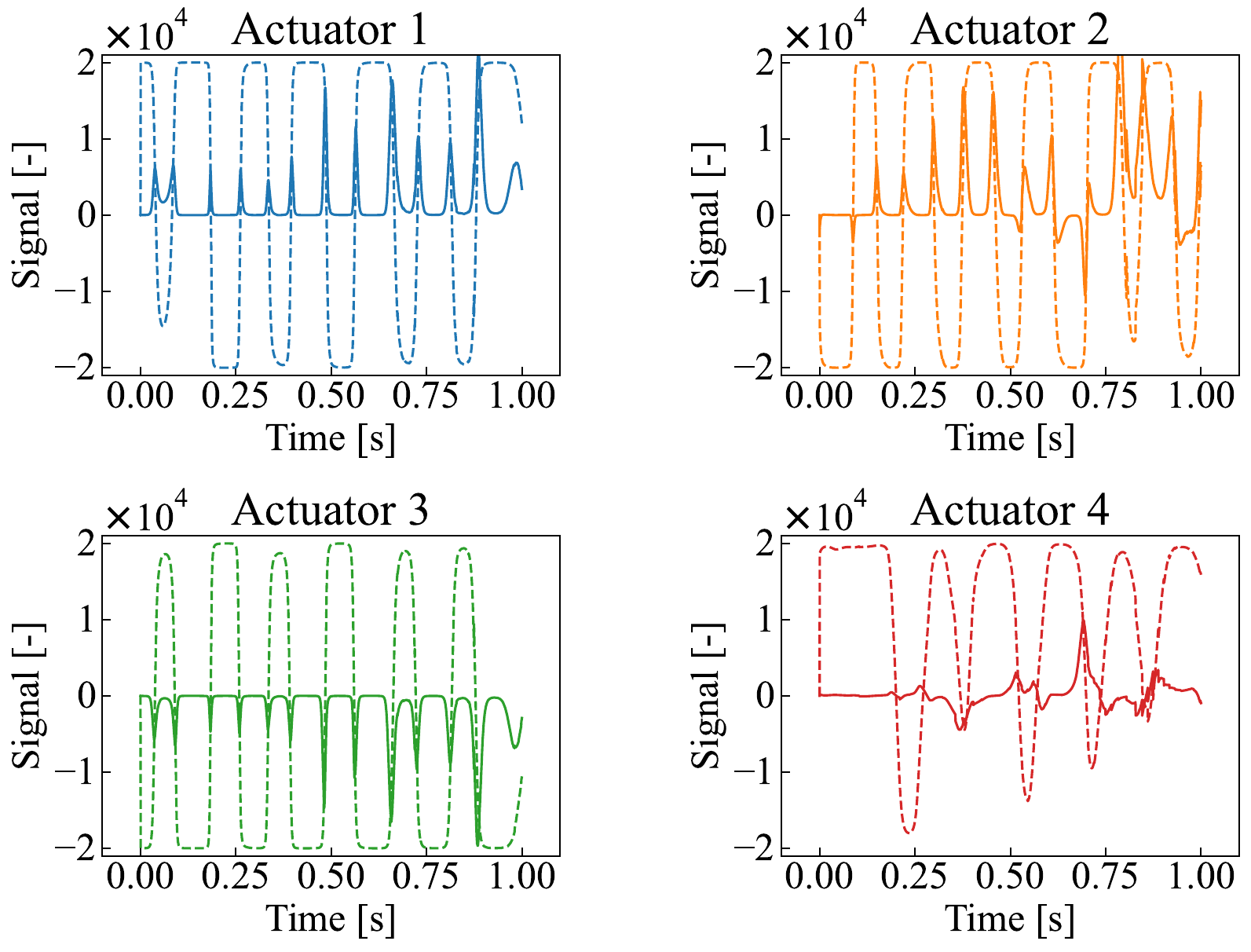}
    \caption{Signals of the respective actuators over the simulation duration. Each line represents the deviation of each actuation signal from its mean value of all test data. Each dashed line represents the actuation signal. Positive and negative actuation signals represent expansion and contraction, respectively.}
    \label{fig:walker_2d_ex1_act_test11_bst}
\end{figure}

To analyze the differences in results by various values of $W$ in more detail, we provide the behaviors of optimized soft robots on two particular test terrains.
Figure~\ref{fig:walker_2d_ex1_test23_wst} shows the behaviors of optimized soft robots on the test terrain where the largest objective function was observed for the soft robot equipped with a feedforward controller (i.e., the soft robot optimized by setting $W=0$).
The red and blue regions respectively represent areas undergoing expansion and contraction through actuation.
The black line represents the terrain on which the orange segment is the range used as the input of the controller.
In Fig.~\ref{fig:walker_2d_ex1_test23_wst}, the soft robot optimized by setting $W=0$ required a time to go over steps during $t=0.35$--$1.0~\mathrm{s}$ while the soft robots obtained by setting $W=0.3$ and $0.4$ advanced over the steps smoothly.
This is because the controller could determine the actuation using the terrain feature in the case of $W=0.3$ and $0.4$.
Figure~\ref{fig:walker_2d_ex1_act_test23_wst} illustrates actuation signals applied to the soft robot optimized by setting $W=0.4$, that is, actuation signals applied to exhibit the dynamics of the rightmost column of Fig.~\ref{fig:walker_2d_ex1_test23_wst}.
Each line represents the deviation of each actuation signal from its mean value of all test data while each dashed line represents the actuation signal.
We observe that the large negative deviation for the actuator 2 and the large positive deviation for the actuator 3 during $t=0.35$--$0.65~\mathrm{s}$, which enables the widening of the gap between the front and rear \textit{legs}, contributing to advancing over the steps.
Figure~\ref{fig:walker_2d_ex1_test11_bst} illustrates the behaviors of optimized soft robots on the test terrain where the smallest objective function was observed for the soft robot equipped with a feedforward controller.
The soft robot optimized by setting $W=0$ lost the balance at a step after $t=0.65~\mathrm{s}$, which improved the objective function by increasing the travel distance.
On the other hand, the soft robots obtained by setting $W=0.3$ and $0.4$ advanced on the step keeping the balance to time $t=1.0~\mathrm{s}$.
Thus, we observed the soft robots equipped with the feedback controller could travel on varying terrains more smoothly.
Figure~\ref{fig:walker_2d_ex1_act_test11_bst} shows actuation signals applied to exhibit the dynamics of the rightmost column of Fig.~\ref{fig:walker_2d_ex1_test11_bst}.
We observe the large positive deviation for the actuator 2 and the large negative deviation for the actuator 3, which implies that the soft robots are not taking larger steps compared to them on other terrains because of the downhill.


\begin{figure}[t]
    \centering
    \includegraphics[width=0.6\textwidth]{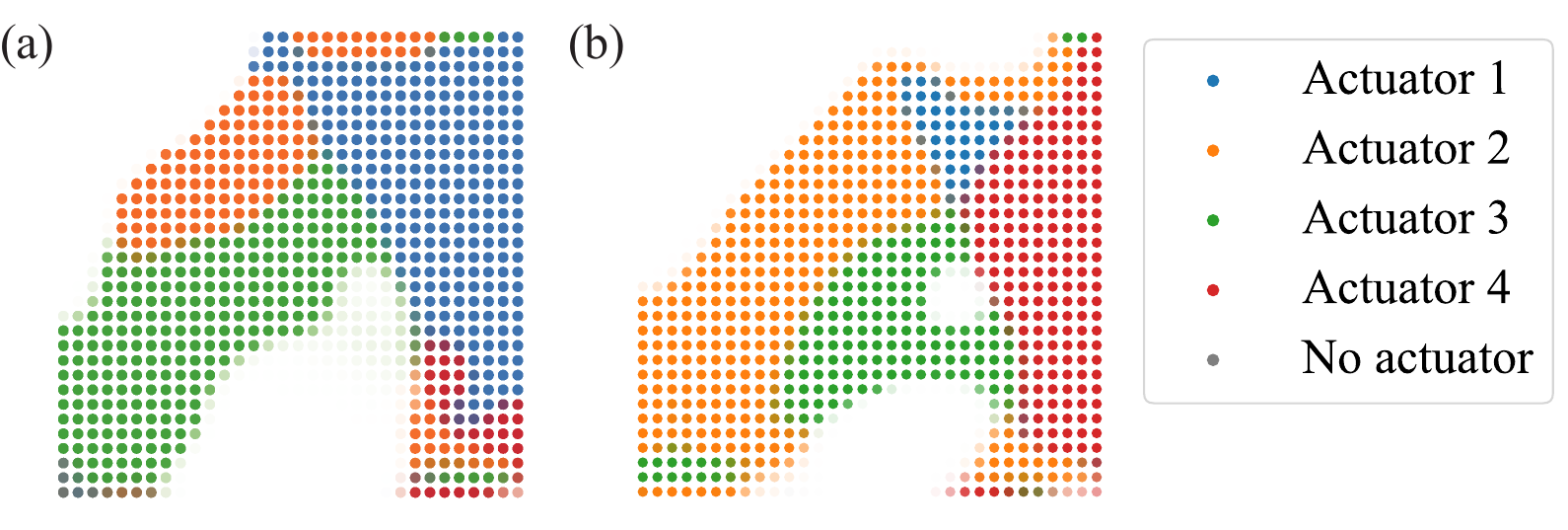}
    \caption{Optimized structures and actuator layouts using training data with the size of (a) $N^\mathrm{data}=16$ and (b) $N^\mathrm{data}=64$.}
    \label{fig:walker_2d_result_ex2}
\end{figure}
\begin{figure}[t]
    \centering
    \includegraphics[width=0.7\textwidth]{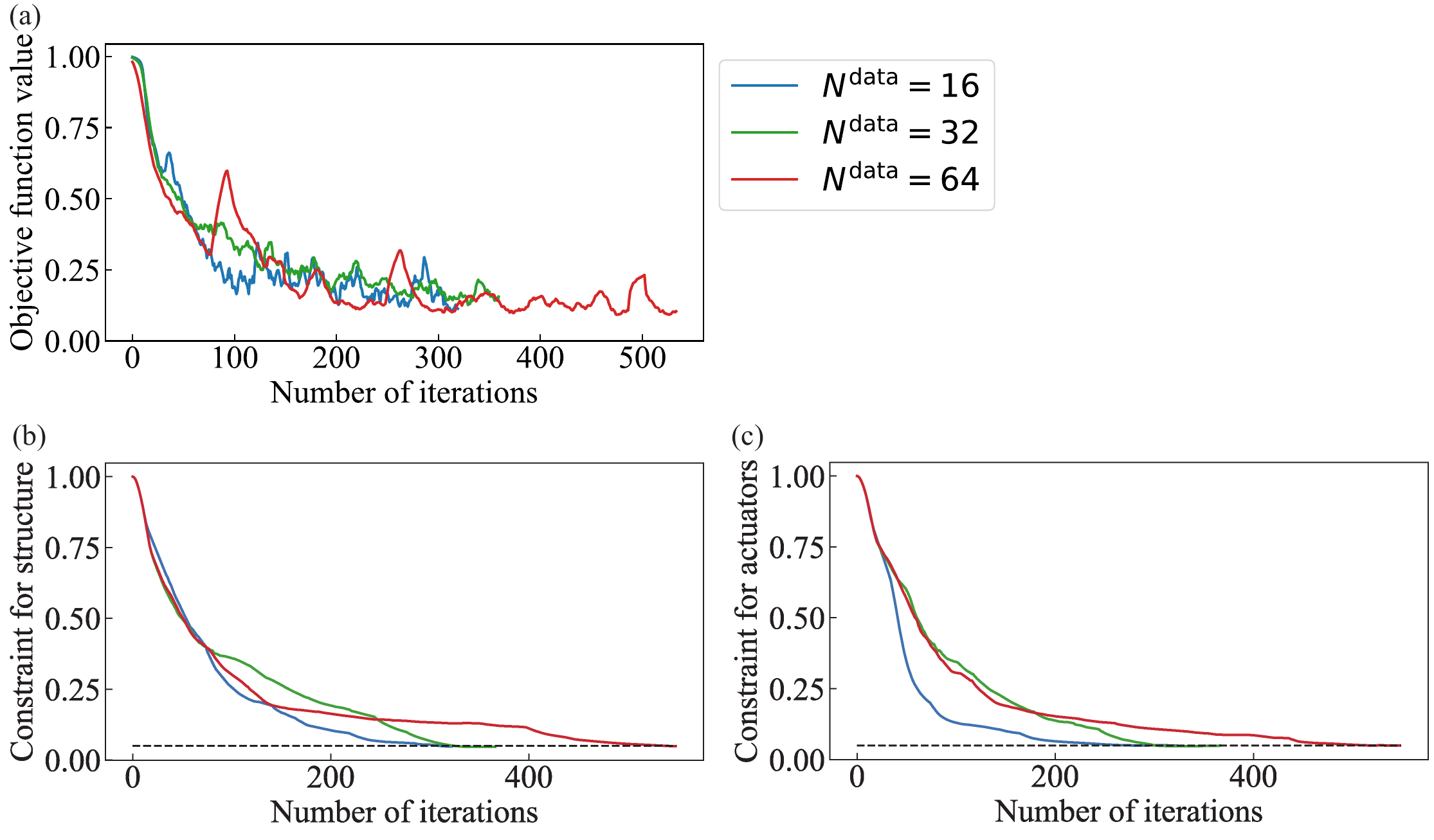}
    \caption{Trajectories of (a) objective function, (b) constraint function for binarization of structure, and (c) constraint function for binarization of actuator layout during optimization with respect to the training dataset size. The objective function values are plotted with a moving average taken every $N^\mathrm{data} / N^\mathrm{batch}$ iterations. Dashed lines in (b) and (c) represents the prescribed upper bounds of the constraint functions.}
    \label{fig:walker_2d_history_ex2}
\end{figure}

\subsubsection{Effect of dataset size}

Next, we examined the effect of training dataset size.
We set the target travel distance as $L=0.7~\mathrm{m}$.
For the size of the surrounding area, we set $W=0.4~\mathrm{m}$.

Figure~\ref{fig:walker_2d_result_ex2} illustrates the optimized structures and actuation layouts obtained using training data with the size of $N^\mathrm{data}=16$ and $64$.
These structures have a common feature of two leg-like substructures as in Fig.~\ref{fig:walker_2d_result_ex1}.
Actuator layouts on the structures also share similarities to the results in Fig.~\ref{fig:walker_2d_result_ex1}; the left and right \textit{legs} are equipped with different actuators.
Especially, the optimized structures in Fig.~\ref{fig:walker_2d_result_ex1}(c) for $N^\mathrm{data}=32$ and Fig.~\ref{fig:walker_2d_result_ex2}(b) for $N^\mathrm{data}=64$ agreed well, which implied that 32 training data was enough to train the walker on one-dimensional terrains.

Figure~\ref{fig:walker_2d_history_ex2} illustrates the history of the objective function $\mathcal{F}$ and constraint functions for binarization of material density and relaxed two-index variables, $\mathcal{C}^\mathrm{to}$ and $\mathcal{C}^\mathrm{lay}$, during the optimization iterations where the objective function was plotted with a moving average taken every $N^\mathrm{data} / N^\mathrm{batch}$ iteration.
We also included the results for $N^\mathrm{data}=32$ in the previous subsection for comparison.
For all cases, the objective function successfully decreased from 1.0 and stayed below 0.25 with oscillation while constraint functions smoothly decreased to satisfy the constraints.
We observed no significant differences among three cases of $N^\mathrm{data}=16$, $32$, and $64$.

\begin{figure}[t]
    \centering
    \includegraphics[width=0.8\textwidth]{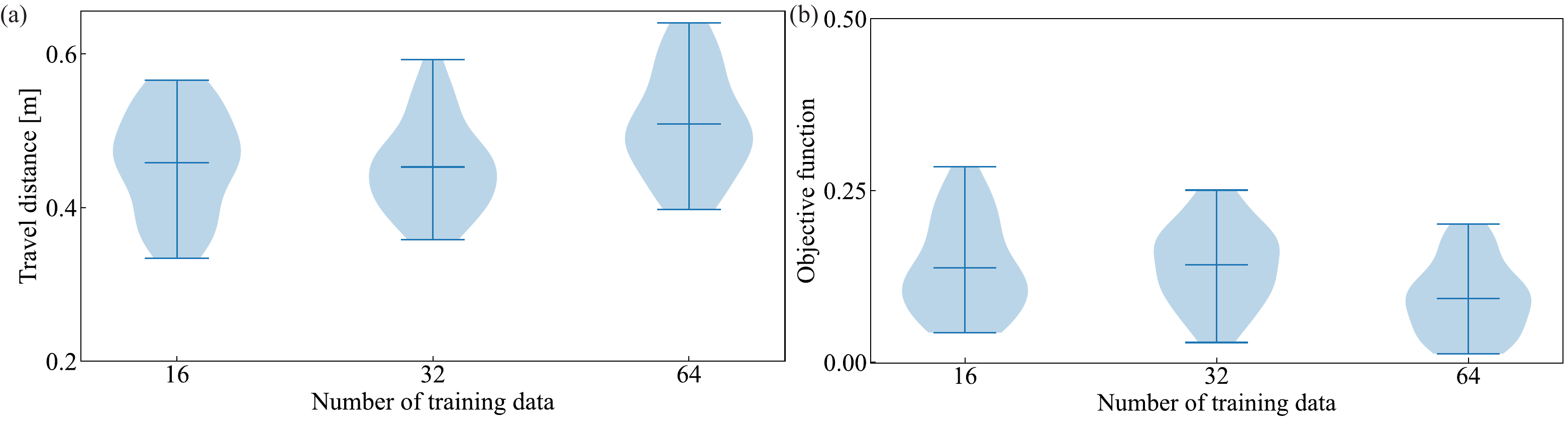}
    \caption{Violin plots illustrating (a) travel distances and (b) objective function values of the optimized soft robots for the test dataset with respect to the training dataset size.}
    \label{fig:walker_2d_violin_ex2}
\end{figure}

Figure~\ref{fig:walker_2d_violin_ex2} shows a violin plot illustrating the travel distances and objective function values evaluated using 32 test data for the soft robots optimized by 16, 32, and 64 training data.
There are no significant differences in travel distances and objective function values in these three cases while large data slightly improved these metrics.
As a result, we confirmed that a few dozen pieces of data suffice to train the walker on one-dimensional terrains.
The objective function values evaluated using the test dataset are almost below 0.25, which we again confirmed that the training process did not result in overfitting.

\subsection{3D walker}
\begin{figure}[t]
    \centering
    \includegraphics[width=0.7\textwidth]{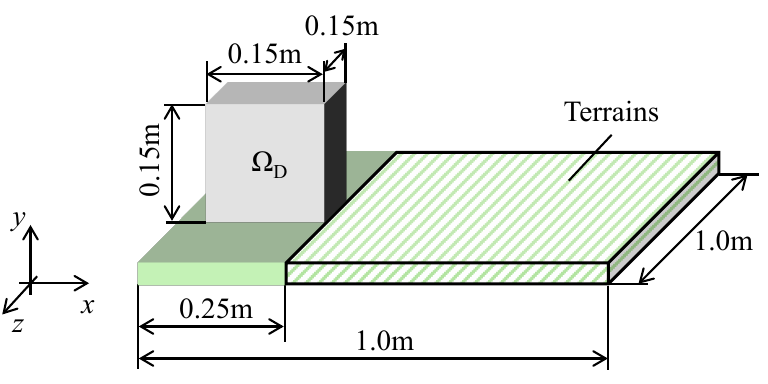}
    \caption{Problem setting for the 3D walker. The gray square represents the design domain of a soft robot and hatched box illustrates an area where terrains are randomly generated. The soft robot is optimized so that the gray square can travel toward the right direction.}
    \label{fig:walker_3d}
\end{figure}
Finally, we designed a soft body that moved toward $x$-direction in three dimensions.
We configured the fixed design domain $\Omega_\mathrm{D}$ as a $0.15 \times 0.15 \times 0.15 \mathrm{m}^3$ cube on a flat ground as shown in Fig.~\ref{fig:walker_3d}.
We set the number of actuators as $N^\mathrm{act} = 6$.
We empirically set $w_\theta=5$ for the weighting coefficient in the objective function \eqref{eq:objective_orig}.
For terrain dataset, we assigned $\varsigma=0.015~\mathrm{m}$ and $\varrho=0.15~\mathrm{m}$, respectively.
We set $c^\mathrm{fb} = 1 / \varsigma = 200/3~\mathrm{m^{-1}}$, the parameter for scaling the feedback signals in Eq.~\eqref{eq:fb_signal}, which normalized the feedback signals independent of the terrain heights.
We set the radius of the surrounding area $W=0.3~\mathrm{m}$, the target travel distance $L=0.7~\mathrm{m}$.
We prepared 64($=N^\mathrm{data}$) training data and also prepared 32 test data.

\begin{figure}[t]
    \centering
    \includegraphics[width=0.7\textwidth]{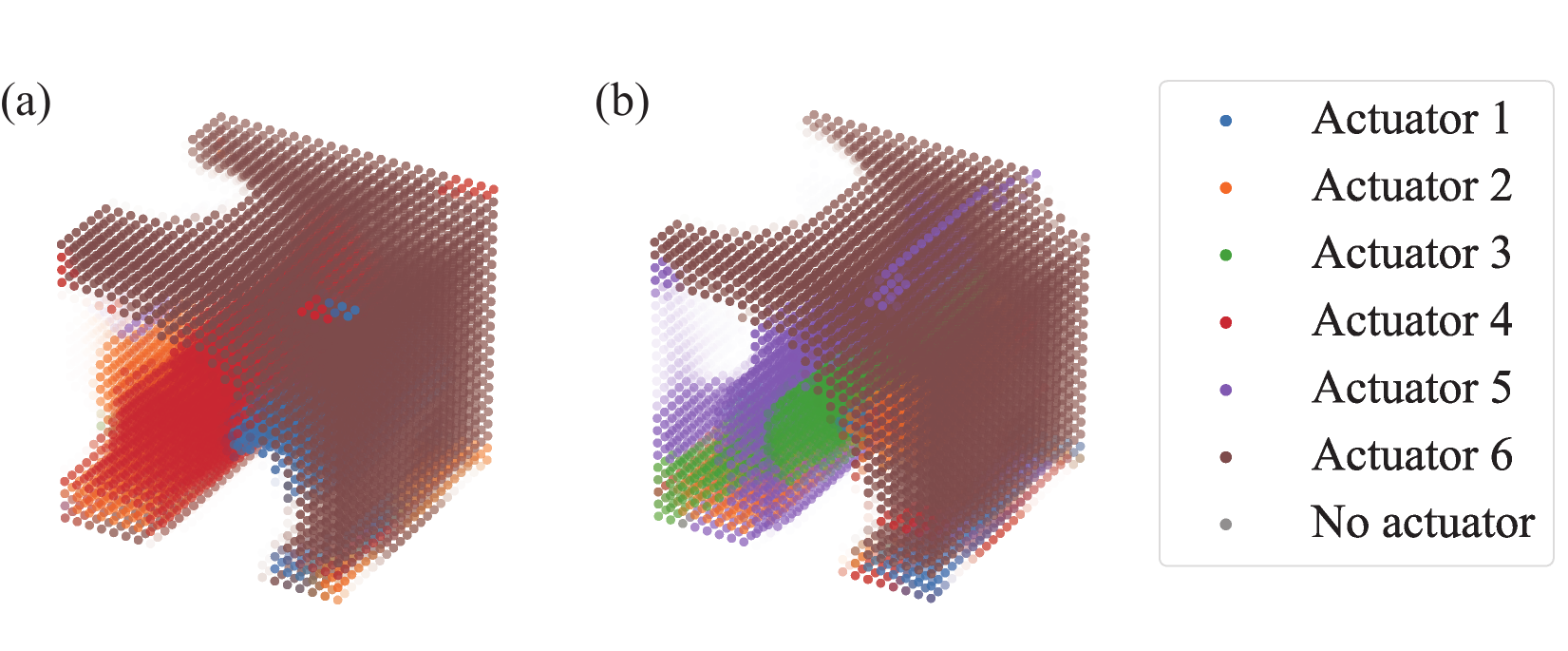}
    \caption{Optimized structures and actuator layouts (a) without the feedback controller and (b) with the feecback controller.}
    \label{fig:walker_3d_result}
\end{figure}

Figure~\ref{fig:walker_3d_result} illustrates the optimized structures and actuation layouts obtained with and without the feedback controller.
These structures resemble each other, that is, two leg-like substructures were formed and they were composed of different actuators.

\begin{figure}[t]
    \centering
    \includegraphics[width=0.7\textwidth]{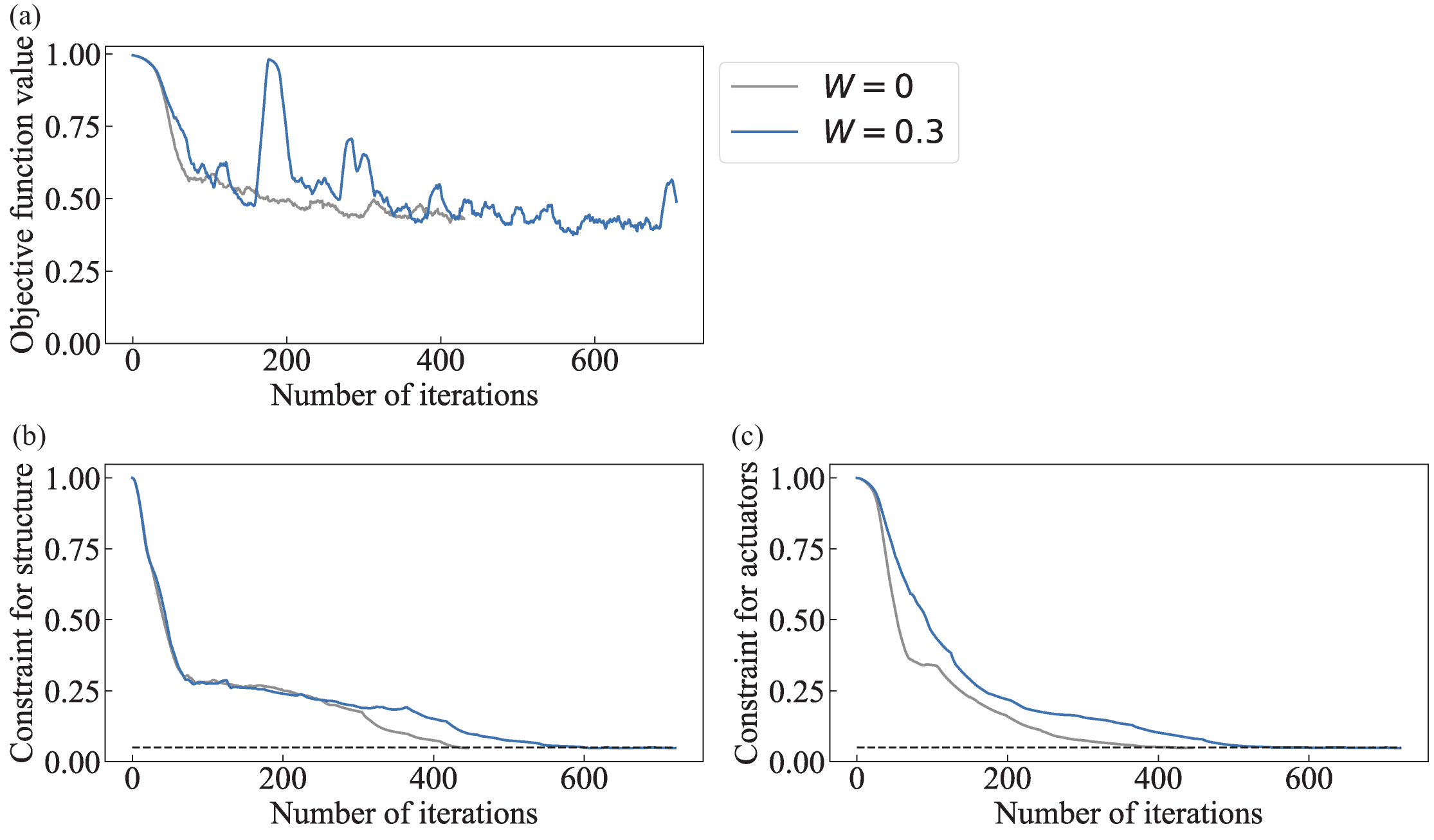}
    \caption{Trajectories of (a) objective function, (b) constraint function for binarization of structure, and (c) constraint function for binarization of actuator layout during optimization. The objective function values are plotted with a moving average taken every $N^\mathrm{data} / N^\mathrm{batch}$ iterations. Dashed lines in (b) and (c) represents the prescribed upper bounds of the constraint functions.}
    \label{fig:walker_3d_history_ex1}
\end{figure}

Figure~\ref{fig:walker_3d_history_ex1} illustrates the history of the objective function $\mathcal{F}$ and constraint functions for binarization of material density and relaxed two-index variables, $\mathcal{C}^\mathrm{to}$ and $\mathcal{C}^\mathrm{lay}$, during the optimization iterations where the objective function was plotted with a moving average taken every $N^\mathrm{data} / N^\mathrm{batch}$ iteration.
For both cases, the objective function successfully decreased from 1.0 to around 0.5 with oscillation while constraint functions smoothly decreased to satisfy the constraints.
These objective function values were larger than those of the two-dimensional problem, which suggests that the three-dimensional problem is more difficult than the two-dimensional one.
This would be because terrains can also vary along with the depth direction and the soft robot requires to keep its posture more complexly.
We would like to explore more sophisticated neural networks for the controller in our future work.

\begin{figure}[t]
    \centering
    \includegraphics[width=0.8\textwidth]{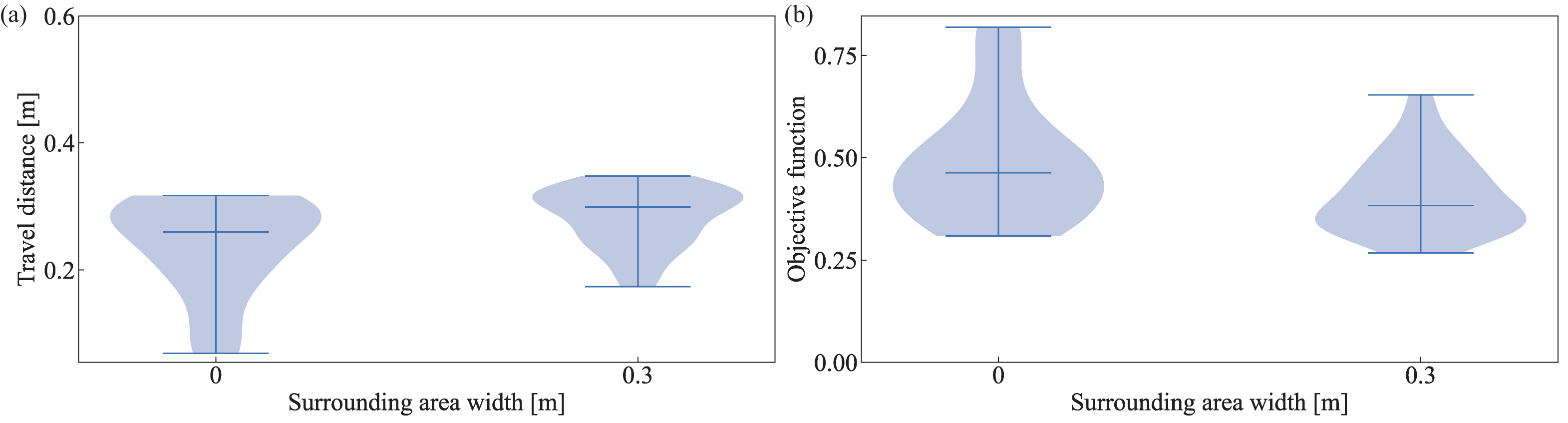}
    \caption{Violin plots illustrating (a) travel distances and (b) objective function values of the optimized soft robots for the test dataset.}
    \label{fig:walker_3d_violin}
\end{figure}

Figure~\ref{fig:walker_3d_violin} shows a violin plot illustrating the travel distances and objective function values evaluated using 32 test data for the soft robots with and without the feedback controller.
While it is slight, the travel distance for $W=0.3$ was larger on average than that for $W=0$ and the objective function for $W=0.3$ resulted in better performance on average than that for $W=0$.
The objective function values evaluated using the test dataset roughly range around 0.5, closely aligning with those obtained from the training dataset, which we again confirmed that the number of training data 64 is still enough for three-dimensional problem.

\begin{figure}[t]
    \centering
    \includegraphics[width=0.7\textwidth]{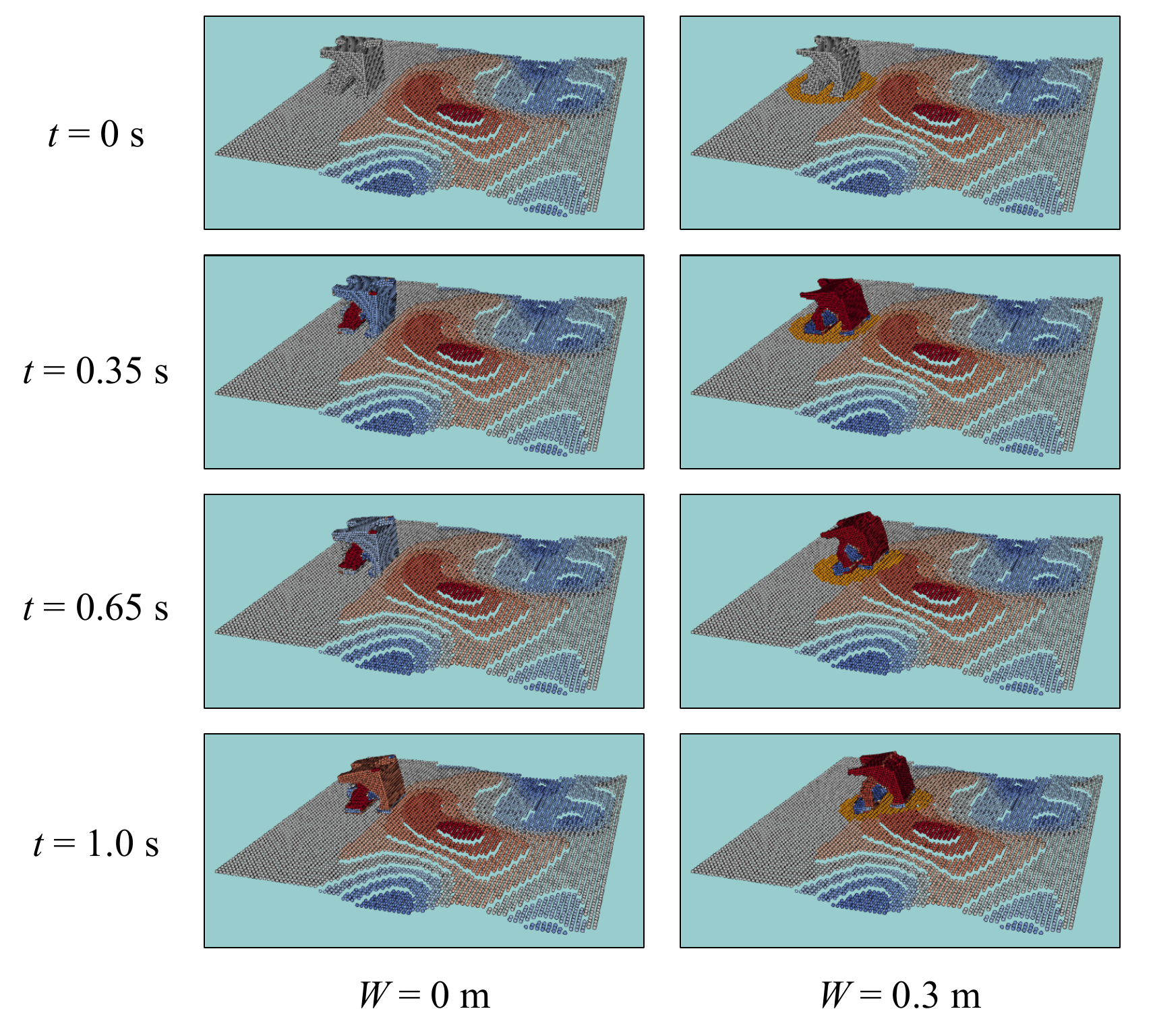}
    \caption{Dynamic behaviors of optimized soft robots for the test terrain on which the largest objective function was observed among the test dataset for the soft robot equipped with a feedforward controller ($W=0$). The red and blue represent the expansion and contraction by actuation, respectively. The orange circle represents the range where the terrain height is inputted to the controller.}
    \label{fig:walker_3d_ex2_test8_wst}
\end{figure}
\begin{figure}[t]
    \centering
    \includegraphics[width=0.7\textwidth]{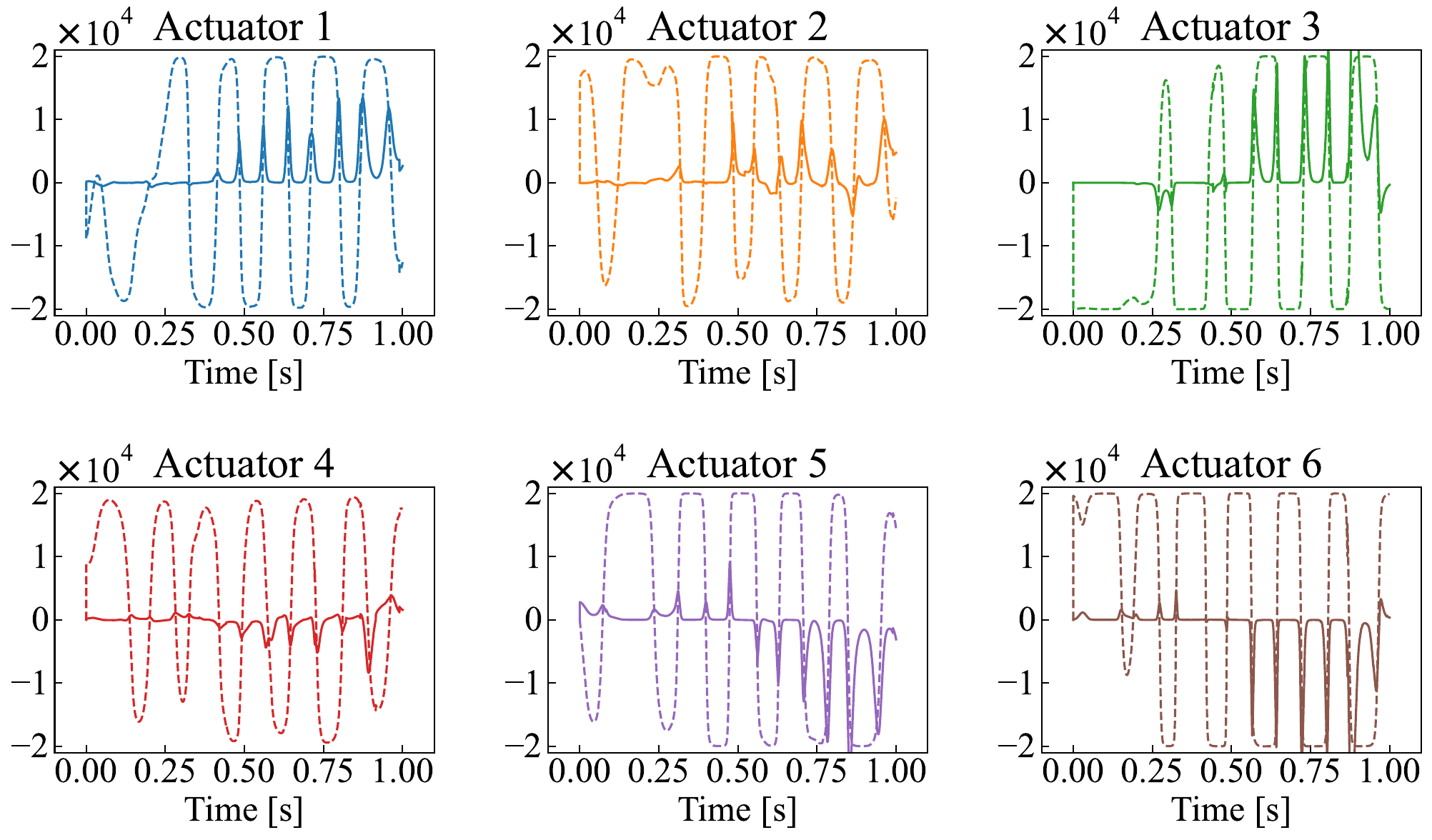}
    \caption{Signals of the respective actuators over the simulation duration. Each line represents the deviation of each actuation signal from its mean value of all test data. Each dashed line represents the actuation signal. Positive and negative actuation signals represent expansion and contraction, respectively.}
    \label{fig:walker_3d_ex2_act_test8_wst}
\end{figure}

\begin{figure}[t]
    \centering
    \includegraphics[width=0.7\textwidth]{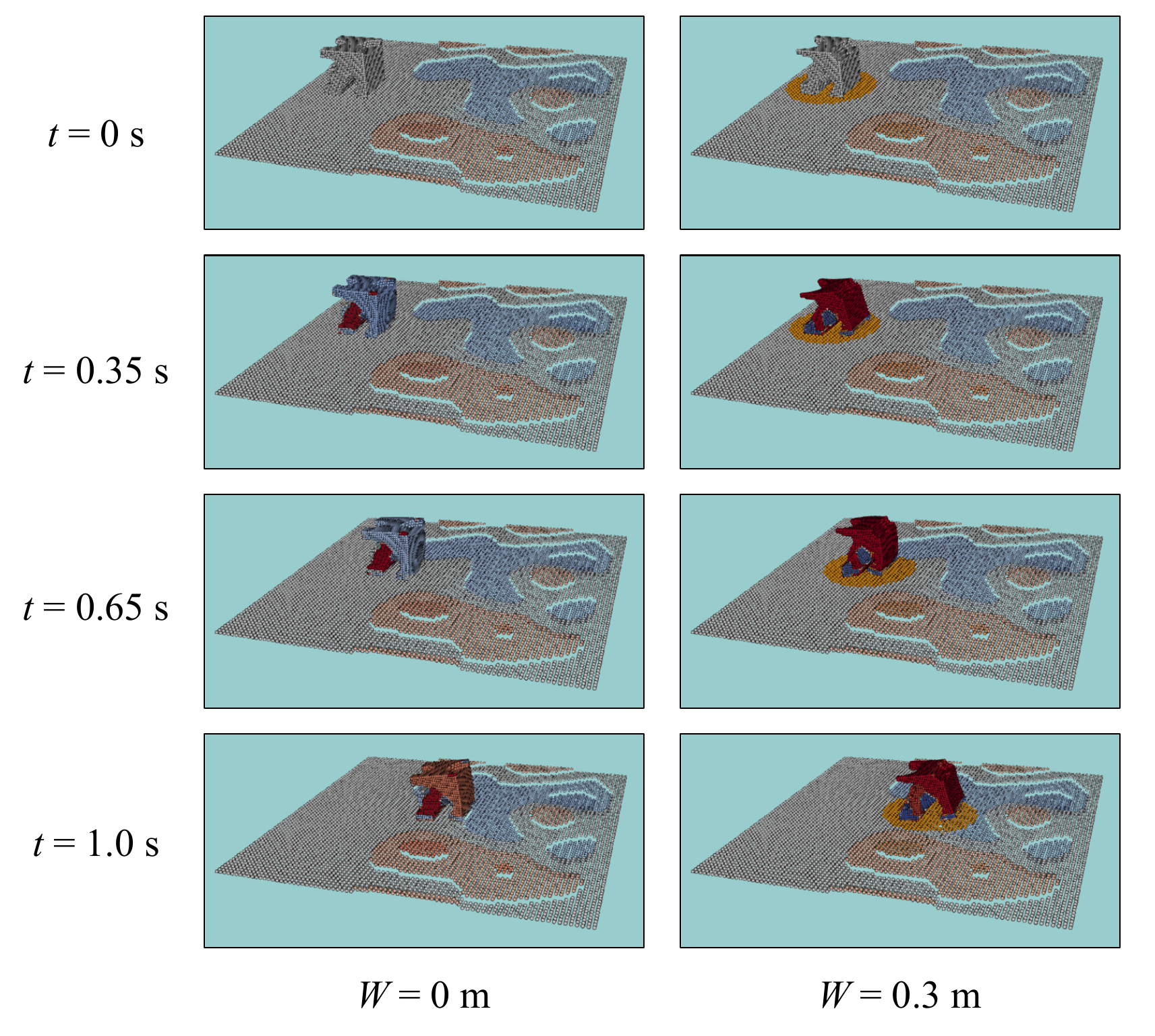}
    \caption{Dynamic behaviors of optimized soft robots for the test terrain on which the smallest objective function was observed among the test dataset for the soft robot equipped with a feedforward controller ($W=0.3$). The red and blue represent the expansion and contraction by actuation, respectively. The orange circle represents the range where the terrain height is inputted to the controller.}
    \label{fig:walker_3d_ex2_test10_bst}
\end{figure}
\begin{figure}[t]
    \centering
    \includegraphics[width=0.7\textwidth]{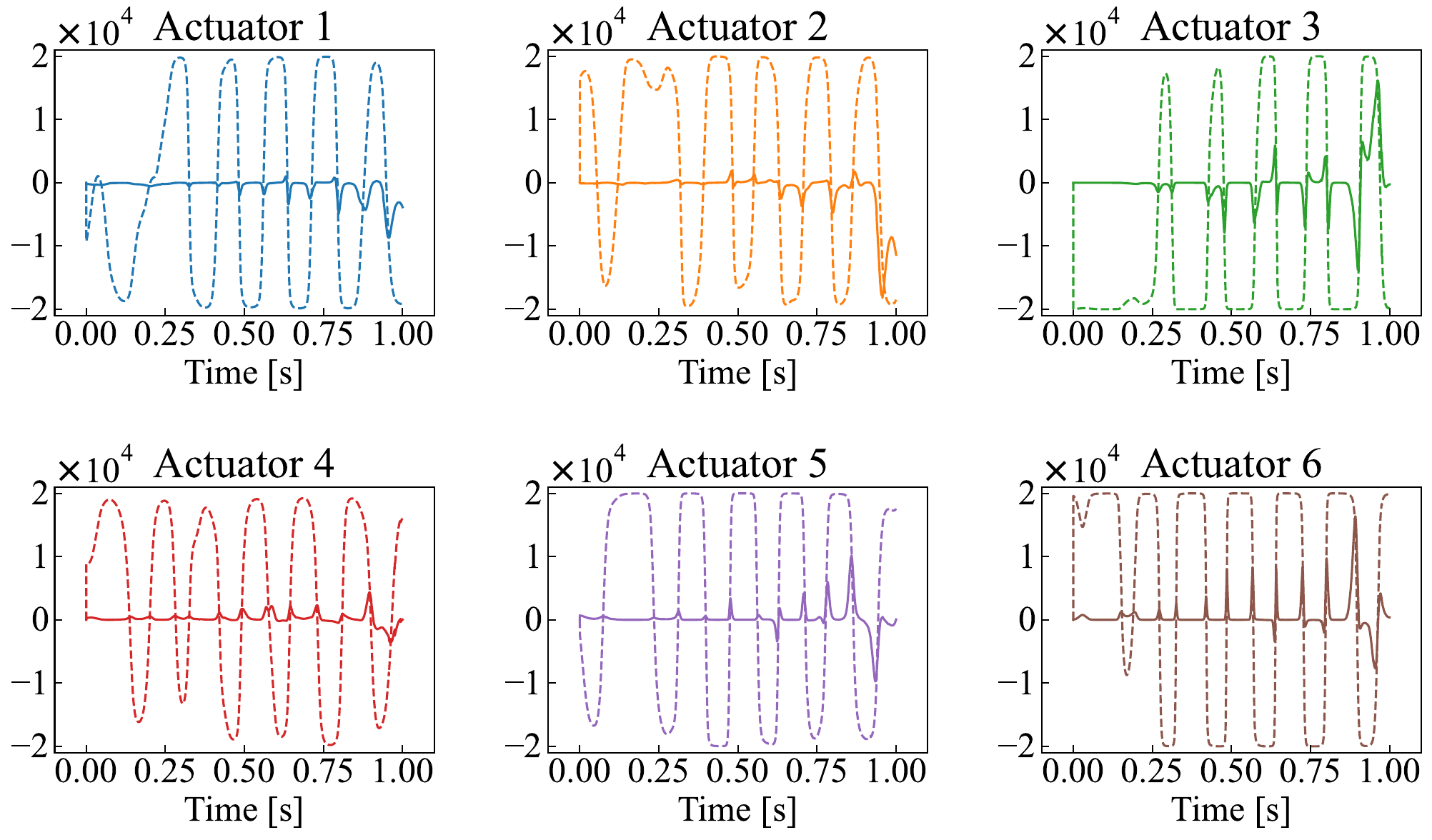}
    \caption{Signals of the respective actuators over the simulation duration. Each line represents the deviation of each actuation signal from its mean value of all test data. Each dashed line represents the actuation signal. Positive and negative actuation signals represent expansion and contraction, respectively.}
    \label{fig:walker_3d_ex2_act_test10_bst}
\end{figure}

To analyze the differences in results with and without the feedback controllers in more detail, we provide the behaviors of optimized soft robots on two particular test terrains.
Figure~\ref{fig:walker_3d_ex2_test8_wst} shows the behaviors of optimized soft robots on the test terrain where the largest objective function was observed for the soft robot equipped with a feedforward controller ($W=0$).
The red and blue regions respectively represent areas undergoing expansion and contraction through actuation.
The orange circle represents the range where the terrain feature is fed into the controller.
In Fig.~\ref{fig:walker_3d_ex2_test8_wst}, the soft robot optimized by setting $W=0$ got stuck in front of the uphill while that obtained by setting $W=0.3$ advanced over the steps smoothly.
Figure~\ref{fig:walker_3d_ex2_act_test8_wst} illustrates actuation signals applied to the soft robot optimized by setting $W=0.3$, that is, actuation signals applied to exhibit the dynamics of the right column of Fig.~\ref{fig:walker_3d_ex2_test8_wst}.
Each line represents the deviation of each actuation signal from its mean value of all test data while each dashed line represents the actuation signal.
The actuators 3 and 5 assigned to the rear \textit{leg} and between \textit{legs} exhibited the opposite tendency, and the deviation of actuator 3 is positive, which contributes to advancing over the uphill.
Figure~\ref{fig:walker_3d_ex2_test10_bst} shows the behaviors of optimized soft robots on the test terrain where the smallest objective function was observed for the soft robot equipped with a feedforward controller.
In Fig.~\ref{fig:walker_3d_ex2_test10_bst}, the terrain in front of the soft robot is flat, so both the soft robots optimized with and without the controller advanced smoothly.
Figure~\ref{fig:walker_3d_ex2_act_test10_bst} shows actuation signals applied to exhibit the dynamics of the rightmost column of Fig.~\ref{fig:walker_3d_ex2_test10_bst}.
We observe that the actuators 3 and 5 assigned to the rear \textit{leg} and between \textit{legs} exhibited the opposite tendency, which contributes to taking a large step forward, but there is no significant deviation because of the flatness of this terrain.
These results again demonstrated the effectiveness of optimization of the feedback controller in addition to the structure.

\section{Conclusions} \label{sec:conclusion}
This paper proposed the computational co-design of soft robots that can travel on varying terrains using the surrounding terrain features.
We employed density-based topology optimization and relaxed multi-indexed optimization for structural and actuator layout optimizations, respectively.
For controller designs, we employed a multilayer perceptron which output the actuation signal from the input of the terrain feature surrounding the soft robot.
We formulated the simultaneous optimization problem of a structure and controller of a soft robot as the locomotion under the uncertainty in terrains and constructed an optimization algorithm, which relied on the stochastic gradient method.
In numerical experiments, we obtained the optimized soft robots having animal-like structures consisting of \textit{legs} and \textit{tendons}, which are adaptively used by the trained controller to advance on various terrains.
We also provided the post-forward analysis for optimized soft robots showcasing that the soft robots equipped with the feedback controller advanced more smoothly on various terrains.

In future work, we would like to integrate more state-of-the-art machine learning techniques, including convolution neural networks for processing environmental features and transformers for processing these time-series features, to handle various tasks beyond locomotion.
We should also consider bridging the gap between simulation and real-world implementation.
This involves the specific actuation model, such as pneumatic, electric, or magnetic actuators, and the manufacturing requirements, i.e., manufacturable structure and actuator layouts.
We will address these issues in our future research and believe that the current study provides fundamental insight toward such a potential direction of co-designing soft robots.

\bibliographystyle{elsarticle-num} 
\bibliography{ref}
\end{document}